\documentclass[12pt,peerreview,draftclsnofoot,letterpaper,oneside,onecolumn]{IEEEtran}

\usepackage{amsfonts,amsmath,amssymb}
\usepackage{cite}
\usepackage{graphicx}
\usepackage{url}
 \newtheorem{theorem}{Theorem}

 \newtheorem{lemma}{Lemma}
  \newtheorem{proposition}{Proposition}
 \newtheorem{corollary}{Corollary}
 \newtheorem{definition}{Definition}
  \newtheorem{remark}{Remark}
  \newtheorem{property}{Property}

\begin{document}

\title{Two Designs of  Space-Time Block Codes Achieving Full Diversity with Partial Interference Cancellation Group Decoding}


\author{Wei~Zhang,~\IEEEmembership{Member,~IEEE,} ~Tianyi~Xu,~\IEEEmembership{Student Member,~IEEE,}
        and ~Xiang-Gen~Xia,~\IEEEmembership{Fellow,~IEEE}
  
 \thanks{W.  Zhang is with School of Electrical Engineering and Telecommunications,   University of New South Wales,
  Sydney, Australia (e-mail: wzhang@ee.unsw.edu.au). His work was supported in part by the Australian Research Council Discovery Project DP1094194.}%
   \thanks{T. Xu and X.-G. Xia are with Department of Electrical and Computer Engineering, University of Delaware, DE 19716, USA (e-mail: \{txu, xxia\}@ee.udel.edu). Their work was supported in part by the Air Force Office of Scientific
Research (AFOSR) under Grant No. FA9550-08-1-0219.
}\thanks{This work was presented in part at
the IEEE Global Telecommunications Conference, Hawaii, USA, December 2009.}
}


%
\maketitle

\begin{abstract}
A partial interference cancellation (PIC) group decoding based space-time block
code (STBC) design criterion was recently proposed by Guo and Xia,
where  the decoding complexity and the code rate  trade-off is dealt
when the full diversity is achieved.
In this paper,  two designs of STBC are proposed for any number of transmit antennas that can obtain full diversity when a PIC group decoding (with
a particular grouping scheme) is applied at receiver.
With the PIC group decoding and an appropriate grouping scheme for the decoding, the proposed STBC are shown to obtain the same diversity gain as the ML decoding, but have a low decoding complexity. The first proposed STBC is designed with multiple diagonal layers and it can obtain the full diversity  for two-layer design   with the PIC group decoding and the rate is up to $2$ symbols per channel use. But with PIC-SIC group decoding, the first proposed STBC can obtain full diversity for any number of layers and the rate can be full.  The second proposed STBC can obtain full diversity and a rate up to  $9/4$ with the PIC group decoding.  Some code design examples  are given and simulation results show that the newly proposed   STBC can well address the rate-performance-complexity tradeoff of the MIMO systems.
\end{abstract}



\section{Introduction}
\label{sec:intro}
Space-time (ST) coding is a
bandwidth-efficient transmission technique that can improve the reliability of data
transmission in MIMO wireless systems \cite{Alamouti,Tarokh00}.
 Orthogonal space-time block coding (OSTBC) is one of the most attractive ST coding approaches
because the special structure of orthogonality guarantees a full diversity and a simple (linear) maximum-likelihood (ML) decoding.
The first OSTBC design was proposed by Alamouti in \cite{Alamouti} for two transmit antennas and was then
extended by Tarokh \emph{et. al.} in \cite{Tarokh00} for any number of transmit antennas.
A class of OSTBC from complex design with the code rate of ${1}/{2}$ was also given by
Tarokh \emph{et. al.} in \cite{Tarokh00}. Later, systematic constructions of complex
OSTBC of rates $(k+1)/(2k)$ for $M=2k-1$ or $M=2k$ transmit antennas for any positive
integer $k$ were proposed in \cite{Liang,SXL,Lu}. However, the OSTBC has a low code rate not more than $3/4$ for more than two transmit antennas  \cite{Wang}.

To enhance the transmission  rate of the STBC, various STBC design approaches were proposed such as quasi-OSTBC \cite{Jafar,Tirkkonen, Papadias,Sharma,SuXia,Aria2,Yuen,Dao,HWang,Karmakar} and algebraic number theory based STBC \cite{Damen,gamal,rajan0,rajan1,kumar1,oggier,kumar2,oggier0,oggier1}. The quasi-OSTBC   increases  the code rate by relaxing the orthogonality condition on the code matrix, which was originally proposed in \cite{Jafar}, \cite{Tirkkonen}, and \cite{Papadias}, independently.
 Due to the group orthogonality, the ML decoding is performed pair-wise or group-wise with an increased complexity compared to the single-symbol decoding.
 In \cite{Khan,HWang,Yuen}, quasi-OSTBC was studied in the sense of minimum decoding complexity, i.e., a real pair-wise symbols decoding. In \cite{Dao,Karmakar,HWang}, the pair-wise decoding was generalized to a general group-wise decoding. The decoding for these codes is the ML decoding and their rates are basically limited by that of OSTBC.
The algebraic number theory based STBC
are designed  mainly based on the ML decoding that may have high complexity and  even though some near-ML decoder, such as sphere decoder
\cite{Viterbo}  can be used, the expected decoding complexity is still dominated by
polynomial terms of a number of symbols which are jointly detected \cite{JaldenSP}.


To reduce the large decoding complexity of the high rate STBC aforementioned, several fast-decodable STBC were recently proposed \cite{Choi}\cite{Hong}. The STBC proposed in \cite{Choi} achieves  a high rate and a reduced decoding complexity at the cost of loss of full diversity. The fast-decodable STBC in \cite{Hong} can obtain   full rate,   full diversity and the reduced ML decoding complexity, but the code design is limited to $2\times 2$ and $4\times 2$ MIMO transmissions only.
Another new perspective of reducing the decoding complexity was recently considered in  \cite{Liu} and \cite{Shang} to resort to conventional linear receivers such as zero-forcing (ZF) receiver or minimum mean square  error (MMSE) receiver  instead of the ML receiver to collect the full diversity. The outage and diversity of linear receivers in flat-fading MIMO channels were studied in \cite{Aria1}, but no explicit code design was given to achieve the full diversity when the linear receivers are used.
Based on the new  STBC design criterion for MIMO systems with linear receivers, Toeplitz STBC \cite{Liu} and overlapped-Alamouti codes \cite{Shang} were proposed and shown to achieve the full diversity with the linear receivers. Recently, some other new designs of STBC with linear receivers were proposed \cite{ZhangICASSP09,ZhangISIT09,WangHM}. However,   the code rate of STBC achieving full diversity with linear receivers  is upper bounded by one. Later, Guo and Xia proposed a partial interference cancellation (PIC) group decoding scheme \cite{Guo} which can be viewed as an intermediate decoding approach between the ML receiver and the ZF receiver by trading a simple single-symbol decoding complexity for a high code rate larger than one symbol per channel use. Moreover, in \cite{Guo} an STBC design criterion was given  to achieve full diversity when the PIC group decoding is applied at the receiver. The proposed PIC group decoding in \cite{Guo} was also connected with the successive interference cancellation (SIC) strategy to aid the decoding process, referred to as PIC-SIC group decoding.  A few code design examples were presented in \cite{Guo}, but a general design of STBC achieving full diversity with the PIC group decoding remains an open problem.

In this paper, we propose two  designs of STBC which can achieve full diversity  with the PIC group decoding for any number of transmit antennas. The first proposed STBC have a structure of multiple diagonal layers  and for each diagonal layer there are exactly $M$ coded symbols embedded, being equal to the number of transmit antennas, which are obtained from a cyclotomic lattice design. Indeed, each diagonal layer of the coded symbols can be viewed as the conventional rate-one diagonal STBC \cite{Damen2,WangGY}. The code rate of the proposed STBC can be  from one to $M$ symbols per channel use by adjusting the codeword length, i.e., embedding different number of layers in the code matrix.
 With the PIC group decoding the  code rate of the first proposed  full-diversity STBC can be only up to $2$ symbols per channel use, i.e., for two layers. For more than two layers embedded in the codeword, the code rate is increased at the cost of losing full diversity with the PIC group decoding. However, with the PIC-SIC group decoding, the proposed STBC with arbitrary number of  layers can obtain full diversity and the code rate can be up to $M$.

 The second proposed STBC is designed with three layers of information symbols embedded in the codeword and the PIC group decoding can be performed in three separate groups accordingly. Without loss of decoding complexity compared to the first proposed STBC, the second proposed STBC
can achieve full diversity and a code rate larger than $2$. Note that the code rate for the first
proposed full-diversity STBC with PIC group decoding can not be above $2$.
In the PIC group decoding of the proposed STBC,   every $M$ neighboring columns of the equivalent channel matrix are clustered into one group.


This paper is organized as follows. A system  model of ST transmission over MIMO channels with the PIC group decoding is introduced in Section \ref{sec:system}.  In Section \ref{sec:new}, a  design of high rate  STBC with the PIC group decoding is proposed, which contains multiple diagonal layers of coded symbols. For a particular code design with two diagonal layers, the full diversity with the PIC group decoding is proved. For the code with PIC-SIC group decoding, the full diversity is   shown for any number of diagonal layers. Several full-diversity code design examples are given in Section \ref{sec:example}.  In Section \ref{sec:xu}, another design of high rate STBC with the PIC group decoding is proposed, which can achieve  full diversity with three layers.    Simulation results are presented in Section \ref{sec:sim}. Finally, in Section \ref{sec:conclusion}, we draw our conclusions.


\emph{Notations}:
  Column vectors (matrices) are denoted by boldface lower
  (upper) case letters. Superscripts $^t$ and $^H$ stand for transpose and
conjugate transpose, respectively. $\mathbb{C}$ denotes the field of complex numbers. $\mathbf{I}_n$ denotes the $n\times n$ identity matrix, and  $\mathbf{0}_{m\times n}$ denotes the $m\times n$ matrix  whose elements are all $0$. $\mathrm{vec}(\mathbf{X})$ is the vectorization of matrix $\mathbf{X}$ by stacking the columns of $\mathbf{X}$ on top each other.

\section{System Model and PIC Group Decoding}
\label{sec:system}

In this section, we first briefly describe the system model
and then describe the PIC group decoding proposed in \cite{Guo}.

\subsection{System Model}

We consider a MIMO transmission  with $M$ transmit antennas and $N$ receive antennas over block fading channels. The received signal matrix $\mathbf{Y}\in\mathbb{C}^{T\times N}$ is
 \begin{eqnarray}\label{eqn:Y}
 \mathbf{Y}=\sqrt{\frac{\rho}{\mu}}\mathbf{X}\mathbf{H}+\mathbf{W},
 \end{eqnarray}
 where $\mathbf{X} \in\mathbb{C}^{T\times M}$ is the   codword matrix,  transmitted over $T$ time slots,  $\mathbf{W}\in\mathbb{C}^{T\times N}$ is a  noise matrix with independent and identically distributed (i.i.d.) entries being circularly symmetric complex Gaussian distributed $\mathcal{CN}(0,1)$, $\mathbf{H}\in\mathbb{C}^{M\times N}$ is the channel matrix whose entries are also i.i.d. with the distribution $\mathcal{CN}(0,1)$, $\rho$ denotes the average signal-to-noise ratio (SNR) per receive antenna and $\mu$ is the normalization factor to ensure that the average  energy of the coded symbols transmitting from all antennas during one symbol period is $1$. The realization of $\mathbf{H}$ is assumed to be known at the receiver, but not known at the transmitter.  Therefore, the signal power is allocated uniformly across the transmit antennas.

\vspace*{3mm}

\begin{definition}[Code Rate] Let $L$ be the number of independent information symbols $\{s_l\}, l=1, \cdots, L$ per codeword $\mathbf{X}$, selected from a complex constellation $\mathcal{A}$. The code rate of the STBC is defined as $R=\frac{L}{T}$ symbols per channel use. If $L=TM$, the STBC is said to have full rate, i.e., $R=M$ symbols per channel use.
\end{definition}
\vspace*{3mm}

In this paper, we consider that  information symbols $\{s_l\}, l=1, \cdots, L$ are coded by linear dispersion STBC as
 \begin{eqnarray}\label{eqn:LSTBC}
 \mathbf{X}  = \sum_{l=1}^L \mathbf{A}_l s_l,
 \end{eqnarray}
where $\mathbf{A}_l\in\mathbb{C}^{T\times M}$ is the linear STBC matrix.

To decode the transmitted sequence $\mathbf{s}$ at the receiver, we need to extract $\mathbf{s}$ from $\mathbf{X}$. This can be done by as follows. By substituting  (\ref{eqn:LSTBC}) into (\ref{eqn:Y}), we get
 \begin{eqnarray}
 \mathbf{Y}&=&\sqrt{\frac{\rho}{\mu}} \sum_{l=1}^L \mathbf{A}_l \mathbf{H}s_l+\mathbf{W}.
 \end{eqnarray}
 Then, by taking vectorization of the matrix $\mathbf{Y}$ we have
\begin{eqnarray}\label{eqn:Y2}
  \mathbf{y}\triangleq \mathrm{vec}(\mathbf{Y})&=&\sqrt{\frac{\rho}{\mu}} \sum_{l=1}^L \mathrm{vec}\left(\mathbf{A}_l \mathbf{H}\right)s_l+\mathrm{vec}(\mathbf{W})\nonumber\\
&=&\sqrt{\frac{\rho}{\mu}}\mathcal{H}\mathbf{s}+\mathbf{w},
\end{eqnarray}
where $\mathbf{y}\in\mathbb{C}^{TN\times 1}$,  $\mathbf{w}\in\mathbb{C}^{TN\times 1}$, $\mathbf{s}=[\begin{array}{cccc}
                                                                                                       s_1 & s_2& \cdots & s_L
                                                                                                    \end{array}]^t$,  and $\mathcal{ {H}}\in\mathbb{C}^{TN\times L}$ is an equivalent channel matrix,
\begin{eqnarray}\label{eqn:Hequi}
\mathcal{ {H}}=\left[\begin{array}{cccc} \mathbf{g}_{1}& \mathbf{g}_{2}& \cdots & \mathbf{g}_{L}\end{array}\right]
\end{eqnarray}
with the $l$-th column  $\mathbf{g}_{l}=\mathrm{vec}\left(\mathbf{A}_l \mathbf{H}\right)$, $l=1,2,\cdots, L$.

For a ZF receiver, the estimate $\mathbf{\hat{s}}^{\mathrm{ZF}}$ of the transmitted symbol sequence $\mathbf{s}$ is,
 \begin{eqnarray}\label{eqn:ZF}
\mathbf{\hat{s}}^{\mathrm{ZF}}=\arg\min_{\mathbf{s}\in\mathcal{A}^L}\left\|
 \mathbf{Q}^{\mathrm{ZF}}\mathbf{y}-\mathbf{s}\right\|^2,
   \end{eqnarray}
 where $\mathbf{Q}^{\mathrm{ZF}}=\sqrt{\frac{\mu}{\rho}}\left(\mathcal{H}^{H}\mathcal{H}\right)^{-1}\mathcal{H}^{H}$. Equivalently, it can be written as the single-symbol decoding as follows,
  \begin{eqnarray}\label{eqn:ZF2}
\mathbf{\hat{s}}_{l}^{\mathrm{ZF}}=\arg\min_{s_l\in\mathcal{A}}\left\|
\left[\mathbf{Q}^{\mathrm{ZF}}\right]_{l,:} \mathbf{y}-s_l\right\|^2,\,\,\,\,l=1,2\cdots, L,
   \end{eqnarray}
    where $\left[\mathbf{Q}^{\mathrm{ZF}}\right]_{l,:}$ denotes the $l$-th row of  $\mathbf{Q}^{\mathrm{ZF}}$.

For an ML receiver, the estimate of $\mathbf{\hat{s}}^{\mathrm{ML}}$   that achieves the minimum of the squared Frobenius norm is given by
 \begin{eqnarray}
\mathbf{\hat{s}}^{\mathrm{ML}}=\arg\min_{\mathbf{s}\in\mathcal{A}^L}\left\|\mathbf{y}-\sqrt{\frac{\rho}{\mu}}\mathcal{H}\mathbf{s}\right\|^2 .
   \end{eqnarray}

In the ML decoding,  computations of squared Frobenius norms for all possible codewords are needed and therefore result in prohibitively huge computational complexity when the length of the  information symbols vector to be decoded is large. In the following, we give a metric to evaluate the computational  complexity of the ML decoding, which is the same as the one shown in \cite[Definition 2]{Hong}.

\vspace*{3mm}

\begin{definition}[Decoding Complexity]
The decoding complexity $\mathcal{O}$ is defined as the number of squared Frobenius norms $\|\cdot\|^2$ that should be computed in the decoding process.
\end{definition}

\vspace*{3mm}

With the above definition, we have the following two remarks.

\vspace*{3mm}
\begin{remark}
The decoding complexity of the ZF detection is $\mathcal{O}=L\cdot|\mathcal{A}|$, i.e., $L$ times of the cardinality of the signal constellation. It is equivalent to the single-symbol decoding complexity.
\end{remark}

\vspace*{3mm}
\begin{remark}
The decoding complexity of the ML detection is $\mathcal{O}=|\mathcal{A}|^L$, i.e., the complexity of the \emph{full} exhaustive search of all $L$ information symbols drawn from the constellation $\mathcal{A}$.
\end{remark}

\vspace*{3mm}
We next describe the PIC group decoding studied in \cite{Guo}.

\subsection{PIC Group Decoding}

Define index set $\mathcal{I}$ as
$$
\mathcal{I}=\{1, 2, \cdots, L\},
$$
where $L$ is the number of information symbols in $\mathbf{s}$. We then partition $\mathcal{I}$ into $P$ groups: $\mathcal{I}_1, \mathcal{I}_2,\cdots, \mathcal{I}_P$ with
$$
\mathcal{I}_p=\{I_{p,1}, I_{p,2},\cdots, I_{p,l_p}\}, \,\,\, p=1,2,\cdots, P,
$$
where $l_p$ is the cardinality of the subset $\mathcal{I}_p$. We call $\mathcal{I}=\{\mathcal{I}_1, \mathcal{I}_2,\cdots, \mathcal{I}_P\}$ a grouping scheme. For such a grouping scheme, we have
$$
\mathcal{I}=\bigcup_{p=1}^{P}\mathcal{I}_p,\,\,\,\,\mathrm{and}\,\,\,\sum_{p=1}^Pl_p=L.
$$
Define
\begin{eqnarray}
\mathbf{s}_{p}&=&\left[\begin{array}{cccc}
                                  s_{I_{p,1}}   & s_{I_{p,2}}  &  \cdots & s_{I_{p,l_p}}
                                 \end{array}
\right]^t,\,\,\,p=1, \cdots, P.\\
\mathbf{G}_{p}&=&\left[\begin{array}{cccc}
                                  \mathbf{g}_{I_{p,1}}   & \mathbf{g}_{I_{p,2}}  &  \cdots & \mathbf{g}_{I_{p,l_p}}
                                 \end{array}
\right],\,\,\,p=1, \cdots, P.
\end{eqnarray}
With these notations, (\ref{eqn:Y2}) can be written as
\begin{eqnarray}
\mathbf{y}=\sqrt{\frac{\rho}{\mu}}\sum_{p=1}^P\mathbf{G}_{p}\mathbf{s}_p+\mathbf{w}.
 \end{eqnarray}

Suppose we want to decode the symbols embedded in the group $\mathbf{s}_p$. The  PIC group decoding first implements linear interference cancellation with a suitable choice of matrix $\mathbf{Q}_p$ in order to completely eliminate the interferences from other groups \cite{Guo}, i.e., $\mathbf{Q}_p\mathbf{G}_q=\mathbf{0}$, $\forall q\neq p$ and $q=1,2,\cdots, P$. Then, we have
\begin{eqnarray}\label{eqn:GZF}
\mathbf{z}_p&\triangleq &\mathbf{Q}_p\mathbf{y}\nonumber\\
&=&\sqrt{\frac{\rho}{\mu}} \mathbf{Q}_p\mathbf{G}_{p}\mathbf{s}_p+\mathbf{Q}_p\mathbf{w},\,\,\,\,\, p=1,2,\cdots, P,
 \end{eqnarray}
where the interference cancellation matrix $\mathbf{Q}_p$ can be chosen as follows \cite{Guo},
\begin{eqnarray}\label{eqn:Qp}
\mathbf{Q}_p=\mathbf{I}_{TN}-\mathbf{G}_p^c \left(\left(\mathbf{G}_p^c\right)^H\mathbf{G}_p^c\right)^{-1}\left(\mathbf{G}_p^c\right)^H,\,\,\,\,\, p=1,2,\cdots, P,
 \end{eqnarray}
in case
\begin{eqnarray}
\mathbf{G}_p^c=\left[\begin{array}{cccccc}
                       \mathbf{G}_1 & \cdots & \mathbf{G}_{p-1} & \mathbf{G}_{p+1} & \cdots & \mathbf{G}_P
                     \end{array}
\right],
 \end{eqnarray}
has full column rank.
  If  $\mathbf{G}_p^c$ does not have
full column rank, then we need to pick a maximal linear independent vector
group from $\mathbf{G}_p^c$ and in this case a projection matrix $\mathbf{Q}_p$ can be
found too \cite{Guo}.

Afterwards, the symbols  in the group $\mathbf{s}_p$ are decoded with the ML decoding algorithm as follows,
 \begin{eqnarray}\label{eqn:GML}
\mathbf{\hat{s}}_p=\arg\min_{\mathbf{s}_p\in\mathcal{A}^{l_p}}\left\|\mathbf{z}_p-\sqrt{\frac{\rho}{\mu}}\mathbf{Q}_p  \mathbf{G}_p\mathbf{s}_p\right\|^2.
   \end{eqnarray}

\vspace*{3mm}

The above PIC group decoding is connected to some of the known
decodings as in the following remarks.

\vspace*{3mm}

\begin{remark}[ML and PIC Group Decoding]
For one special case of $P=1$, the grouping scheme is $\mathcal{I}=\{\mathcal{I}_1\}$ with $\mathcal{I}_1=\mathcal{I}$. From (\ref{eqn:Qp}), we have $\mathbf{Q}_p=\mathbf{I}_{TN}$. Then,  the PIC group decoding is equivalent to the ML decoding where   all information symbols are jointly decoded.
\end{remark}
\vspace*{3mm}

\begin{remark}[ZF and PIC Group Decoding]
For the special case of $P=L$, the grouping scheme is
$
\mathcal{I}=\{\mathcal{I}_1,\mathcal{I}_2,\cdots, \mathcal{I}_L\}=\{\{1\}, \{2\}, \cdots, \{L\}\},
$  i.e., every single symbol is regarded as one group. Then,  the  PIC  group decoding is equivalent to the ZF decoding where every single symbol is separated from all the other symbols and then decoded.
\end{remark}
\vspace*{3mm}

\begin{remark}[ZF, ML and PIC Group Decoding]
The PIC group decoding with $1\leq P\leq L$ can be viewed as an intermediate decoding approach  between the ML decoding and the ZF decoding. Alternatively, the ML decoding and the ZF decoding can both be regarded as the special cases of the PIC group decoding corresponding to $P=1$ and $P=L$, respectively.
\end{remark}
\vspace*{3mm}

\begin{remark}[PIC Group Decoding Complexity] For the PIC group decoding, the following two steps are needed: the group zero-forcing to cancel the interferences coming from all the other groups as shown in (\ref{eqn:GZF}) and the  group ML decoding to jointly decode the symbols in one group  as shown in (\ref{eqn:GML}). Therefore, the decoding complexity of the PIC group decoding should reside in the above two steps. Note that   the interference cancellation process shown in (\ref{eqn:GZF})  mainly involves with linear matrix computations, whose computational complexity  is small compared to the ML decoding for an exhaustive search of all candidate symbols. Therefore, to evaluate the decoding complexity of the PIC group decoding, we mainly focus on the computational complexity of the ML decoding within the PIC group decoding algorithm. According to \emph{Definition 2}, the ML decoding complexity in the PIC group decoding algorithm is   $\mathcal{O}= \sum_{p=1}^P |\mathcal{A}|^{l_p}$. It can be seen that the PIC group decoding provides a flexible decoding complexity which can be from the ZF decoding complexity $L|\mathcal{A}|$ to the ML decoding complexity $|\mathcal{A}|^{L}$.
\end{remark}

\begin{remark}[PIC-SIC Group Decoding] In \cite{Guo}, an SIC-aided PIC group decoding algorithm, namely PIC-SIC group decoding was proposed. Similar to the  BLAST detection algorithm \cite{Foschini}, the PIC-SIC group decoding is performed  after removing the already-decoded symbol set from the received signals to reduce the interference. If each group has only one symbol, then the PIC-SIC group will be equivalent to the BLAST detection.

\end{remark}

\subsection{STBC Design Criterion with PIC Group Decoding}

The performance of a decoding algorithm
for a wireless communication system is related to the diversity order. If
the average probability of a detection error for communication over a fading channel usually behaves as:
$$
P_e(\mathrm{SNR})\leq  c \cdot \mathrm{SNR}^{-G_d}
$$
where $c$ is a constant and $G_d$ is called the \emph{diversity order} of the system. For an MIMO communication system, the maximum diversity order is $M N$, i.e., the product of the number of transmit antennas and the number of receiver antennas. In order to optimize the reception performance of the MIMO system, a full diversity is usually pursued which can be achieved by   a proper signal transmission scheme or data format (e.g., STBC). In \cite{Tarokh00}, the ``rank-and-determinant criterion'' of STBC design was proposed to maximize both the diversity gain $G_d$ and the coding gain $\frac{1}{c}$ of the MIMO system with an ML decoding.  Recently, in \cite{Guo} an STBC design criterion was derived to achieve full diversity when the PIC group decoding is used at the receiver. In the following, we cite the main result of the STBC design criterion proposed in \cite{Guo}.

\vspace*{3mm}

\begin{proposition}\label{prop1}
\cite[Theorem 1]{Guo} [\emph{Full-Diversity Criterion under PIC Group Decoding}]

For an  STBC  $\mathbf{X}$  with the PIC group decoding, the full diversity is achieved when
  \begin{enumerate}
    \item   the code $\mathbf{X}$ satisfies the full rank criterion, i.e.,
it achieves full diversity when
the ML receiver is used; \emph{and}
    \item    $\mathbf{G}_1, \mathbf{G}_2,\cdots, \mathbf{G}_P$    are linearly independent vector groups  for any $\mathbf{H}\neq \mathbf{0}$.
  \end{enumerate}
\end{proposition}

 \vspace*{3mm}

In \cite{Guo},  the STBC achieving full diversity with PIC group decoding were proposed for $2$ and $4$ transmit antennas. However,  a systematic code design of the full-diversity STBC with PIC group decoding remains an open problem. 

\vspace*{3mm}

\begin{proposition}\label{prop2}
\cite[p.4374]{Guo} [\emph{Full-Diversity Criterion under PIC-SIC Group Decoding}]

For an  STBC  $\mathbf{X}$  with the PIC-SIC group decoding, the full diversity is achieved when
  \begin{enumerate}
    \item   the code $\mathbf{X}$ satisfies the full rank criterion, i.e.,
it achieves full diversity when
the ML receiver is used; \emph{and}
    \item   at each decoding stage, $\mathbf{G}_{q_1}$, which corresponds to the current to-be decoded symbol group $\mathbf{s}_{q_1}$, the remaining groups $\mathbf{G}_{q_2},  \cdots, \mathbf{G}_{q_L}$  corresponding to yet uncoded symbol groups    are linearly independent vector groups  for any $\mathbf{H}\neq \mathbf{0}$.
  \end{enumerate}
\end{proposition}

\section{A Design of STBC with PIC Group Decoding}\label{sec:new}
In this section, we first propose a systematic design of high-rate STBC  which has a rate   up to $M$ symbols per channel use and achieves full diversity with the ML decoding. The systematic design of the STBC is structured with multiple diagonal layers.  Then, we prove that the proposed STBC with  two diagonal layers can obtain full diversity with the PIC group decoding and the code rate can be up to $2$ symbols per channel use. Finally, we prove that the proposed STBC with any number of diagonal layers can obtain full diversity with PIC-SIC group decoding and the code  rate can be up to $M$ symbols per channel use.

 \subsection{Encoding Technique}
Our proposed space-time code $\mathbf{C}$,  i.e., $\mathbf{X}$ in (\ref{eqn:Y}), is of size $T\times M$ (for any given $T$, $M$ and $T\geq M$) and will be transmitted from $M$ antennas over $T$ time slots. Let $P=T-M+1$. The symbol stream  $\{s_l\}, l=1, \cdots, L$  (composed of $L=MP$ complex symbols chosen from QAM constellation and then scaled by $1/\sqrt{E[|s_l|^2]}$) is first parsed into $M\times 1$ symbol vectors $\mathbf{s}_p$ ($p=1,2,\cdots,P$). Each symbol vector is linearly precoded by an $M\times M$ matrix $\mathbf{\Theta}$, which is a chosen constellation rotation matrix.   Next, the $M\times 1$ vector $\mathbf{\Theta}\mathbf{s}_p$ is used to form the space-time code matrix $\mathbf{C}$, in which the $p$-th descending diagonal from left to right is the diagonal form of $\mathbf{\Theta}\mathbf{s}_p$.

The resulting transmitted code matrix $\mathbf{C}$ is given by
\begin{eqnarray}\label{eqn:code}
\mathbf{C} = \left[\begin{array}{cccc}
                     X_{1,1} & 0  & \cdots  &  0 \\
                    X_{2,1}& X_{1,2} & \ddots  & \vdots  \\
                    \vdots & X_{2,2} &  \ddots &  0 \\
                    X_{P,1} & \vdots & \ddots & X_{1,M}  \\
                     0 & X_{P,2} & \ddots  & X_{2,M} \\
                     \vdots & 0  & \ddots  & \vdots \\
                     0 &  \vdots &  \ddots & X_{P,M}
                  \end{array}
 \right],
 \end{eqnarray}
 where the $p$-th descending diagonal from left to right, denoted by $\mathbf{X}_p =\left[\begin{array}{cccc}
                      X_{p,1}   & X_{p,2}& \cdots &  X_{p,M}
                    \end{array}
  \right]^t$  is given by
  \begin{eqnarray}\label{eqn:RM}
  \mathbf{X}_p =\mathbf{\Theta}\mathbf{s}_p, \,\,\,\,\,p=1,2,\cdots, P
   \end{eqnarray}
and the $M\times 1$ information symbol vector $\mathbf{s}_p$ is given by
  \begin{eqnarray}
  \mathbf{s}_p =\left[\begin{array}{cccc}
                      s_{(p-1)M+1}   & s_{(p-1)M+2}& \cdots &  s_{pM}
                    \end{array}
  \right]^t,
   \end{eqnarray}
   for $ \,\,\,\,\,p=1,2,\cdots, P$.

\vspace*{3mm}

\begin{proposition}
The proposed STBC in (\ref{eqn:code}) has asymptotically full rate when the block length is sufficiently large.
\end{proposition}

\vspace*{1mm}
\begin{proof} In the codeword in (\ref{eqn:code}), a total number of $MP$ independent information symbols are encoded into the codeword $\mathbf{C}$, which is then transmitted from $M$ antennas over $T$ time slots. The code rate of transmission is therefore
   \begin{eqnarray}\label{eqn:rate}
   R=\frac{MP}{T}=\frac{M(T-M+1)}{T}=M\left(1-\frac{M-1}{T}\right).
   \end{eqnarray}
   For a very large block length $T$, it can be seen that the rate $R$ of the proposed ST coding scheme  approaches $M$ symbols per channel use, i.e. the full rate.
\end{proof}

\subsection{Choice of   Rotation Matrix $\mathbf{\Theta}$}

In \cite{WangGY}, the rotation matrix $\mathbf{\Theta}$ was designed for diagonal STBC to achieve the full diversity gain and the optimal diversity product. With the optimal cyclotomic lattices design for $M$ transmit antennas, from \cite[Table I]{WangGY} we can get a set of integers $(m,n)$ and let $K=mn$. Then, the optimal lattice $\mathbf{\Theta}$ is given by  \cite[Eq. (16)]{WangGY}
  \begin{eqnarray}\label{eqn:cyclo}
 \mathbf{\Theta}=\left[\begin{array}{llll}
                          \zeta_K & \zeta_K^2 & \cdots & \zeta_K^M \\
                         \zeta_{K}^{1+n_2m} & \zeta_{K}^{2(1+n_2m)} & \cdots & \zeta_{K}^{M(1+n_2m)}\\
                         \vdots & \vdots & \ddots & \vdots \\
                       \zeta_{K}^{1+n_Mm} & \zeta_{K}^{2(1+n_Mm)} & \cdots & \zeta_{K}^{M(1+n_Mm)}
                       \end{array}
 \right].
 \end{eqnarray}
 where
$\zeta_K=\exp(\mathbf{j}2\pi/K)$
with $\mathbf{j}=\sqrt{-1}$ and $n_2,n_3, \cdots, n_M$ are distinct integers such that $1+n_im$ and $K$ are co-prime for any $2\leq i \leq M$.

\vspace*{3mm}

\emph{Example 1}:  For $4$ transmit antennas we can choose $m=3, n=5$ and $K=15$ according to \cite[Table I]{WangGY}. Then, in order to ensure that  $1+n_im$ and $K$ are co-prime for any $2\leq i \leq M$  we can obtain $n_2=1$, $n_3=2$, $n_4=4$. When $m=3$, the signal constellation is located on the equal literal
triangular lattice. When $m=4$, $n$ can be $4$ and
$n_i$ can be $0,1,2,3$, and in this case the signal constellation is located on the square lattice.
 %
\vspace*{3mm}

\emph{Example 2}:  For $5$ transmit antennas we can select $m=n=5$ and $K=25$. Then, $n_2=1$, $n_3=2$, $n_4=3$, $n_5=4$.

\vspace*{3mm}

The cyclotomic design of the matrix $\mathbf{\Theta}$ is vital for the design of the algebraic STBC. In the following, we show some properties of the matrix $\mathbf{\Theta}$ that will be used later for our design.

\vspace*{3mm}

\begin{property} \cite{WangGY} 
 The diagonal cyclotomic ST code $\Omega$   defined by $\Omega=\left\{\mathrm{diag}\left[ \begin{array}{cccc}
                                                    X_1 & X_2 & \cdots & X_M
                                                  \end{array}
   \right]\right\}$
  achieves full diversity under ML decoding, where
   $\left[ \begin{array}{cccc}
                                                    X_1 & X_2 & \cdots & X_M
                                                  \end{array}
   \right]^t=\mathbf{\Theta }\left[\begin{array}{cccc}
                                                    s_1 & s_2 & \cdots & s_M
                                                  \end{array}
  \right]^t$ and $\mathbf{\Theta}$ is given by (\ref{eqn:cyclo}).
\end{property}
\vspace*{3mm}

\begin{property}
Every entry of the matrix $\mathbf{\Theta}$ in (\ref{eqn:cyclo}) is non-zero.
\end{property}

This property is obvious from (\ref{eqn:cyclo}).

\subsection{Achieving Full Diversity with ML Decoding}

We show the main result of the proposed STBC when an ML decoding is used at the receiver, as follows.
\vspace*{3mm}

\begin{theorem}[Full Diversity with ML Decoding]
Consider a MIMO transmission with $M$ transmit antennas and $N$ receive antennas over block fading channels. The  STBC $\mathbf{C}$ as described  in (\ref{eqn:code}) achieves full diversity under the ML decoding.
\end{theorem}

\vspace*{1mm}
\begin{proof}[Proof of Theorem 1]
In order to prove  that the ST code   $\mathbf{C}$ in (\ref{eqn:code}) can obtain full diversity under ML decoding, it is sufficient to prove that $\Delta_\mathbf{C}=\mathbf{C}-\mathbf{\hat{C}}$ achieves full rank for any distinct pair of ST codewords $\mathbf{C}$ and $\mathbf{\hat{C}}$.

   For any   pair of distinct codewords $\mathbf{C}$ and $\mathbf{\hat{C}}$, there exists at least one index $p$ ($1\leq p\leq P$) such that $\mathbf{X}_p-\mathbf{\hat{X}}_p\neq \mathbf{0}$, where $\mathbf{X}_p$ and $\mathbf{\hat{X}}_p$ are related to $\mathbf{s}_p$ and $\mathbf{\hat{s}}_p$ from (\ref{eqn:RM}), respectively. Let $p$ denote  the minimum index of vectors satisfying $\mathbf{X}_p-\mathbf{\hat{X}}_p\neq \mathbf{0}$. Then, for any index $q$ with $q<p$, it must have $\mathbf{X}_q-\mathbf{\hat{X}}_q= \mathbf{0}$. Define $\breve{X}=X-\hat{X}$ as the difference between symbols $X$ and $\hat{X}$. Then, from (\ref{eqn:code}) $\Delta_\mathbf{C}$  can be expressed as
   \begin{eqnarray}
\Delta_\mathbf{C} = \left[\begin{array}{cccc}
                        0 & 0  & \cdots  &  0 \\
                         \vdots & \vdots  & \ddots  &  \vdots \\
                     0 & \vdots & \ddots  &  \vdots \\
                    \breve{X}_{p,1}&  0 & \ddots  & \vdots  \\
                    \vdots & \breve{X}_{p,2} &  \ddots &  \vdots \\
                    \breve{X}_{P,1} & \vdots & \ddots &  0  \\
                     0 & \breve{X}_{P,2} & \ddots  & \breve{X}_{p,M} \\
                     \vdots & 0  & \ddots  & \vdots \\
                     0 &  \vdots &  \ddots & \breve{X}_{P,M}
                  \end{array}
 \right],
 \end{eqnarray}
   where   $\breve{X}_{p,m}\neq 0$ for $m=1,2,\cdots, M$. This is because for $\mathbf{X}_p-\mathbf{\hat{X}}_p\neq \mathbf{0}$, it exists $\mathbf{s}_p-\mathbf{\hat{s}}_p\neq \mathbf{0}$. Due to the suitably chosen constellation rotation matrix $\mathbf{\Theta}$ in (\ref{eqn:cyclo}),  $\mathbf{X}_p-\mathbf{\hat{X}}_p$ must have nonzero entries for any $\mathbf{s}_p\neq\mathbf{\hat{s}}_p$. Then, the matrix $\Delta_\mathbf{C}$ has full rank.

   The full rankness of  $\Delta_\mathbf{C}$ can be examined
similar to that for the Toeplitz code (or delay diversity code) \cite{Liu}  by checking if the columns of $\Delta_\mathbf{C}$ are linearly independent. Specifically, we establish  $\Delta_\mathbf{C}\bar{\alpha}=\mathbf{0}$ with
   $\bar{\alpha}=[\begin{array}{cccc}
                                                                                                               \alpha_1 & \alpha_2 & \cdots & \alpha_M
                                                                                                             \end{array}
   ]^t$. First, we examine the $p$-th equation in $\Delta_\mathbf{C}\bar{\alpha}=\mathbf{0}$ and   get $\alpha_1 \breve{X}_{p,1}=0$. Because  $\breve{X}_{p,m}\neq 0$ for $m=1,2,\cdots, M$, $\alpha_1=0$. Then, we examine the $(p+1)$-th equation and get $\alpha_2 \breve{X}_{p,2}=0$. Immediately, $\alpha_2=0$. Likewise, we  examine the $(p+2)$-th equation until $(p+M-1)$-th equation, and we can get $\alpha_1=\alpha_2=\cdots=\alpha_M=0$. Therefore, all columns of $\Delta_\mathbf{C}$ are linearly independent and  $\Delta_\mathbf{C}$  has full rank.
\end{proof}

This property will be used in next section in the  proof of the full diversity property
under the PIC group decoding.

\subsection{Achieving Full Diversity with PIC Group Decoding when $P=2$}

We show the main result of the proposed STBC when a PIC group decoding with a particular grouping scheme is used at the receiver, as follows.

\vspace*{3mm}

\begin{theorem}[Full Diversity with PIC Group Decoding when $P=2$] Consider a MIMO transmission with $M$ transmit antennas and $N$ receive antennas  over block fading channels. The STBC $\mathbf{C}$ as described in (\ref{eqn:code}) with two diagonal layers (i.e., $P=2$) is used at the transmitter. The equivalent channel matrix is $\mathcal{H}\in \mathbb{C}^{TN\times MP}$.  If the received signal is decoded using the PIC group decoding with the grouping scheme $\mathcal{I}=\{\mathcal{I}_1, \mathcal{I}_2\}$ where $
 \mathcal{I}_p=  \{(p-1)M+1, (p-1)M+2, \cdots, pM  \}$ for $p=1,2$, i.e., the size of each group is equal to the number of transmit antennas $M$, then the code $\mathbf{C}$ achieves the full diversity. The code rate of the full-diversity STBC can be up to $2$ symbols per channel use.
\end{theorem}

 \vspace*{1mm}

In order to prove \emph{Theorem 2}, let us first introduce the following lemmas.

 \vspace*{3mm}

\begin{lemma}
Consider the system as described in \emph{Theorem 2} with $N=1$  and the STBC $\mathbf{C}$  as given by (\ref{eqn:code}),
\begin{enumerate}
  \item    the   equivalent channel matrix  $\mathcal{H}\in \mathbb{C}^{{T\times MP}}$ can be expressed as
    \begin{eqnarray}\label{eqn:H0}
   \mathcal{H}=\left[\begin{array}{cccc}
                       \mathbf{G}_1  &  \mathbf{G}_2 &  \cdots &   \mathbf{G}_P
                     \end{array}
   \right]
  \end{eqnarray}
  where
   \begin{eqnarray}\label{eqn:G}
   \mathbf{G}_p= \left[\begin{array}{c}
                        \mathbf{0}_{(p-1)\times M} \\
                        \mathrm{diag}(\mathbf{h})\mathbf{\Theta} \\
                        \mathbf{0} _{(P-p)\times M}
                      \end{array}
 \right],\,\,\,\, p=1,2, \cdots, P,
   \end{eqnarray}
  \item When $P=2$,  $\mathbf{G}_1$ and $\mathbf{G}_2$    are linearly independent vector groups as long as $\mathbf{h}\neq \mathbf{0}$, where
$\mathbf{h}=\mathbf{H}$.
\end{enumerate}
 \end{lemma}

\vspace*{1mm}

 A proof of \emph{Lemma 1} is given in Appendix I.

 \vspace*{3mm}

\begin{lemma}
Consider the system as described in \emph{Theorem 2} and the STBC $\mathbf{C}$  as given by (\ref{eqn:code}). For the equivalent channel matrix $\mathcal{H}\in \mathbb{C}^{{TN\times MP}}$, $\mathbf{G}_1, \mathbf{G}_2,\cdots, \mathbf{G}_P$    are linearly independent vector groups for $\mathbf{h}\neq \mathbf{0}$ when $N>1$ if and only if $\mathbf{G}_1, \mathbf{G}_2,\cdots, \mathbf{G}_P$    are linearly independent vector groups for $\mathbf{h}\neq \mathbf{0}$ when $N=1$.
 \end{lemma}
 \vspace*{1mm}
The proof of \emph{Lemma 2} is straightforward and is also the same as
what is mentioned  in \cite{Guo}.

\vspace*{3mm}

\begin{proof}[Proof of Theorem 2]
  As shown in \emph{Proposition \ref{prop1}},    a  codeword  $\mathbf{C}$  with PIC group decoding can obtain the full diversity if
   \begin{description}
     \item[1)]   $\mathbf{C}$  achieves the full diversity with the ML receiver, and
     \item[2)] $\mathbf{G}_1, \mathbf{G}_2,\cdots, \mathbf{G}_P$    are linearly independent vector groups as long as $\mathbf{h}\neq \mathbf{0}$.
   \end{description}

For the proposed code $\mathbf{C}$ in (\ref{eqn:code}) with $P=2$, the first condition is satisfied as shown in \emph{Theorem 1}.  The second condition is satisfied as shown in \emph{Lemma 1} for $N=1$ and \emph{Lemma 2} for $N>1$, respectively. Therefore,  the  code $\mathbf{C}$ in (\ref{eqn:code}) with $P=2$ can obtain full diversity with the PIC group decoding provided that the grouping scheme  is $\mathcal{I}=\{\mathcal{I}_1, \mathcal{I}_2\}$ where $
 \mathcal{I}_p=  \{(p-1)M+1, (p-1)M+2, \cdots, pM  \}$ for $p=1,2$.

The code rate of the full-diversity STBC with the PIC group decoding can be derived by substituting $P=2$ and $T=M+1$ into (\ref{eqn:rate}) as
$$
R=\frac{2M}{M+1}.
$$
For a large number of transmit antennas, the rate approaches to (but not larger than) $2$ symbols per channel use.
\end{proof}
 \vspace*{3mm}

\begin{corollary}
The decoding complexity of the PIC group decoding of the proposed STBC with the grouping scheme as described in \emph{Theorem 2} is $\mathcal{O}=P\cdot|\mathcal{A}|^M$.
\end{corollary}

\vspace*{3mm}

\begin{remark}
The decoding complexity of the proposed STBC with the PIC group decoding is equivalent to the ML decoding of $M$ independent information symbols jointly. As shown in (\ref{eqn:rate}), the code rate of the proposed STBC in (\ref{eqn:code}) for a given $M$ can be increased by embedding larger number of groups in the codeword, i.e., increasing the value of $P$. It is noteworthy to mention that the increase of the code rate does not result in the increase of the decoding complexity.
\end{remark}

\subsection{Achieving Full Diversity with PIC-SIC Group Decoding}

For the proposed STBC with any number of layers and the PIC-SIC group decoding we have the following results.

\vspace*{3mm}

\begin{theorem}[Full Diversity with PIC-SIC Group Decoding] Consider a MIMO transmission with $M$ transmit antennas and $N$ receive antennas  over block fading channels. The STBC $\mathbf{C}$ as described in (\ref{eqn:code}) with $P$ diagonal layers  is used at the transmitter. The equivalent channel matrix is $\mathcal{H}\in \mathbb{C}^{TN\times MP}$.  If the received signal is decoded using the PIC-SIC group decoding with the sequential order  and  with the grouping scheme being $\mathcal{I}=\{\mathcal{I}_1, \cdots, \mathcal{I}_P\}$ where $\mathcal{I}_p=  \{(p-1)M+1, (p-1)M+2, \cdots, pM  \}$ for $p=1, \cdots, P$, i.e., the size of each group is equal to the number of transmit antennas $M$, then the code $\mathbf{C}$ achieves the full diversity. The code rate of the full-diversity STBC can be up to $M$ symbols per channel use.
\end{theorem}

 \vspace*{1mm}

The proof of this theorem is simple. Observing that  $\mathcal{H}=\left[\begin{array}{cccc}
                       \mathbf{G}_1  &  \cdots &     \mathbf{G}_P
                     \end{array}
   \right]$ and the group $\mathbf{G}_p$ is linearly independent from the groups $\{\mathbf{G}_{p+1}, \cdots, \mathbf{G}_P\}$ for any $p$ and $1\leq p\leq P-1$ where $\mathbf{G}_{p}$ is given by (\ref{eqn:G}), according to \emph{Proposition 2} the full diversity can be easily proved.
The detailed proof is omitted.
 \vspace*{3mm}

   \section{Code Design Examples}\label{sec:example}
 In this section, we show a few code design examples. We denote $\mathbf{C}_{M,T,P}$ the code constructed by (\ref{eqn:code}) for given parameters: $M$ the number of transmit antennas, $T$ the block length of the code, and $P$ the number of groups to be decoded in the PIC group decoding. 
 For notational brevity,  we only show the equivalent channel of the proposed codes for MISO systems.

\subsection{For Two Transmit Antennas}
Consider a code for 2 transmit antennas with 3 time slots. According to the code structure (\ref{eqn:code}), we have
\begin{eqnarray}
\mathbf{C}_{2,3,2}=\left[\begin{array}{cc}
                   X_{1,1}  &   \\
                   X_{2,1}  & X_{1,2}   \\
                     &  X_{2,2}
                 \end{array}
\right],
\end{eqnarray}
where $[\begin{array}{cc}
           X_{1,1}  &   X_{1,2}
        \end{array}
]^t=\mathbf{\Theta}[\begin{array}{cc}
          s_1 &   s_2
        \end{array}
]^t$ and $[\begin{array}{cc}
           X_{2,1}  &   X_{2,2}
        \end{array}
]^t=\mathbf{\Theta}[\begin{array}{cc}
          s_3 &   s_4
        \end{array}
]^t$. The constellation rotation matrix $\mathbf{\Theta}$ can be chosen as
$$
\mathbf{\Theta}=\left[\begin{array}{rr}
                        \gamma  &  \delta  \\
                          -\delta  & \gamma
                      \end{array}
\right],
$$ where $\gamma=\cos \theta$ and $\delta=\sin \theta$ with $\theta=1.02$ \cite{Guo}.

The code rate of the code is $4/3$. In fact, this code is equivalent to the one proposed in \cite[Section VI - \emph{Example 1}]{Guo}.

The equivalent channel of the code $\mathbf{C}_{2,3,2}$ is given by
\begin{eqnarray}
\mathcal{H}=\left[\begin{array}{cccc}
                    \gamma h_1  & \delta h_1  &   &   \\
                      -\delta h_2 & \gamma h_2  &  \gamma h_1  & \delta h_1    \\
                      &  &    -\delta h_2 & \gamma h_2
                  \end{array}
\right].
 \end{eqnarray}
The grouping scheme for the PIC  group decoding is $\mathcal{I}_1=\{1,2\}$ and $\mathcal{I}_2=\{3,4\}$. It can be seen that $\mathbf{G}_1$
and $\mathbf{G}_2$ are linearly independent. Then, the code can obtain full diversity with the PIC group decoding.

\subsection{For Four Transmit Antennas}

For given $T=5$, the code achieving full diversity with the PIC group decoding can be designed as follows,
\begin{eqnarray}\label{eqn:C452}
\mathbf{C}_{4,5,2} = \left[\begin{array}{cccc}
                     X_{1,1} &    &    &   \\
                    X_{2,1}& X_{1,2} &    &   \\
                      & X_{2,2} &  X_{1,3} &   \\
                      &   & X_{2,3} & X_{1,4}  \\
                       &   &    & X_{2,4}
                  \end{array}
 \right].
 \end{eqnarray}
This code has a code rate of $8/5$ and two groups to be decoded.  The equivalent channel of the code $\mathbf{C}_{4,5,2}$ is
 \begin{eqnarray}
\mathcal{H}=\left[ \begin{array}{ccc}
                    \left[\begin{array}{c}
                         \mathbf{B} \\
                         \mathbf{0}_{1\times 4}
                       \end{array}
    \right]   & \left[\begin{array}{c}
                          \mathbf{0}_{1\times 4} \\
                         \mathbf{B}
                       \end{array}
    \right]
                   \end{array}
\right],
 \end{eqnarray}
 where $\mathbf{B}$ is given by
 \begin{eqnarray}\label{eqn:B}
 \mathbf{B}=\left[\begin{array}{cccc}
                      h_1\theta_{1,1} & h_1\theta_{1,2}  &  h_1\theta_{1,3} &  h_1\theta_{1,4}     \\
                       h_2\theta_{2,1} & h_2\theta_{2,2}  &  h_2\theta_{2,3} &  h_2\theta_{2,4}    \\
                         h_3\theta_{3,1} & h_3\theta_{3,2}  &  h_3\theta_{3,3} &  h_3\theta_{3,4}   \\
                          h_4\theta_{4,1} & h_4\theta_{4,2}  &  h_4\theta_{4,3} &  h_4\theta_{4,4}
               \end{array} \right]
 \end{eqnarray}
with $\theta_{i,j}$ being  the ($i,j$)-th entry of the matrix $\mathbf{\Theta}$ for $i,j=1,2, 3, 4$.
 The grouping scheme for  the PIC group decoding is $\mathcal{I}_1=\{1,2, 3, 4\}$ and  $\mathcal{I}_2=\{5,6,7,8\}$. It can be seen that the groups $\mathbf{G}_1$  and $\mathbf{G}_2$ are linearly independent to each other. Then, the code can obtain full diversity with the PIC group decoding.

 Consider $T=6$ time slots.   We can get
\begin{eqnarray}\label{eqn:C462}
\mathbf{C}_{4,6,2} = \left[\begin{array}{cccc}
                     X_{1,1} &    &    &   \\
                     0 & X_{1,2} &    &   \\
                    X_{2,1} & 0  &  X_{1,3} &   \\
                      & X_{2,2} & 0  & X_{1,4}  \\
                       &   & X_{2,3}  & 0  \\
                      &    &   &  X_{2,4}
                  \end{array}
 \right].
 \end{eqnarray}
 The code rate of the code $\mathbf{C}_{4,6,2}$ is $4/3$ which has the same rate as the one proposed in \cite[Section VI - \emph{Example 2}]{Guo}.  The equivalent channel of the code $\mathbf{C}_{4,6,2}$ is
 \begin{eqnarray}
\mathcal{H}=\left[ \begin{array}{ccc}
                    \left[\begin{array}{c}
                         \mathbf{B} \\
                         \mathbf{0}_{2\times 4}
                       \end{array}
    \right]   & \left[\begin{array}{c}
                          \mathbf{0}_{2\times 4} \\
                         \mathbf{B}
                       \end{array}
    \right]
                   \end{array}
\right],
 \end{eqnarray}
 where $\mathbf{B}$ is given by (\ref{eqn:B}).
Because the groups $\mathbf{G}_1$  and $\mathbf{G}_2$ are linearly independent to each other. Then, the code can obtain full diversity with the PIC group decoding.

Moreover, we can also design the code for $T=6$ with $3$ layers (i.e., $P=3$) as follows,
\begin{eqnarray}\label{eqn:C463}
\mathbf{C}_{4,6,3} = \left[\begin{array}{cccc}
                     X_{1,1} &    &    &   \\
                    X_{2,1}& X_{1,2} &    &   \\
                    X_{3,1} & X_{2,2} &  X_{1,3} &   \\
                      & X_{3,2} & X_{2,3} & X_{1,4}  \\
                       &   & X_{3,3}  & X_{2,4} \\
                      &    &   &  X_{3,4}
                  \end{array}
 \right].
 \end{eqnarray}
The code rate of the code $\mathbf{C}_{4,6,3}$ is $2$  and the equivalent channel is given by
\begin{eqnarray}
\mathcal{H}=\left[ \begin{array}{ccc}
                    \mathbf{G}_1 & \mathbf{G}_2 & \mathbf{G}_3
                   \end{array}
\right],
 \end{eqnarray}
 where
 \begin{eqnarray}
 \begin{array}{ccc}
    \mathbf{G}_1=\left[\begin{array}{c}
                         \mathbf{B} \\
                         \mathbf{0}_{1\times 4} \\
                          \mathbf{0}_{1\times 4}
                       \end{array}
    \right], &
    \mathbf{G}_2=\left[\begin{array}{c}
                         \mathbf{0}_{1\times 4}\\
                         \mathbf{B} \\
                          \mathbf{0}_{1\times 4}
                       \end{array}
    \right], &
     \mathbf{G}_3=\left[\begin{array}{c}
                          \mathbf{0}_{1\times 4} \\
                         \mathbf{0}_{1\times 4}\\
                         \mathbf{B} \\
                       \end{array}
    \right].
 \end{array}\nonumber
 \end{eqnarray}

It can be proved that the groups $\mathbf{G}_1$, $\mathbf{G}_2$, and $\mathbf{G}_3$ are not linearly independent groups. Therefore, according to \emph{Proposition 1}, the code $\mathbf{C}_{4,6,3}$ cannot achieve the full diversity with the PIC group decoding.

However, the code $\mathbf{C}_{4,6,3}$ can obtain full diversity with PIC-SIC group decoding. This is because $\mathbf{G}_1$ is linearly independent from $\mathbf{G}_2$ and $\mathbf{G}_3$, and $\mathbf{G}_2$ is linearly independent from $\mathbf{G}_3$. According to  \emph{Proposition 2},  with PIC-SIC group decoding and a proper decoding order $\{1,2,3\}$ or $\{3,2,1\}$, the code $\mathbf{C}_{4,6,3}$ can achieve the full diversity.

\subsection{For Five Transmit Antennas}
For given $T=6$ and $P=2$, the code is designed as follows,
\begin{eqnarray}\label{eqn:C5}
\mathbf{C}_{5,6,2} = \left[\begin{array}{ccccc}
                     X_{1,1} &    &    &   & \\
                    X_{2,1}& X_{1,2} &    & &  \\
                      & X_{2,2} &  X_{1,3} & &  \\
                      &   & X_{2,3} & X_{1,4} & \\
                       &   &    & X_{2,4}& X_{1,5}\\
                       & & & & X_{2,5}
                  \end{array}
 \right].
 \end{eqnarray}
The code rate of the code $\mathbf{C}_{5,6,2}$ is $5/3$.  The equivalent channel is
\begin{eqnarray}
\mathcal{H}=\left[ \begin{array}{ccc}
                    \left[\begin{array}{c}
                         \mathrm{diag}(\mathbf{h})\mathbf{\Theta}_5 \\
                         \mathbf{0}_{1\times 5}
                       \end{array}
    \right]   & \left[\begin{array}{c}
                          \mathbf{0}_{1\times 5} \\
                          \mathrm{diag}(\mathbf{h})\mathbf{\Theta}_5
                       \end{array}
    \right]
                   \end{array}
\right],
 \end{eqnarray}
 where $\mathbf{\Theta}_5$ is the rotation matrix of size $5\times 5$.

The grouping scheme for  the PIC group decoding of $\mathbf{C}_{5,6,2}$ is $\mathcal{I}_1=\{1,2, 3, 4, 5\}$ and $\mathcal{I}_2=\{6,7,8,9,10\}$. It can be seen that the groups $\mathbf{G}_1$ and $\mathbf{G}_2$ are linearly independent to each other. Then, the code $\mathbf{C}_{5,6,2}$ can obtain full diversity with the PIC group decoding.


\section{Another Design of STBC with PIC Group Decoding}\label{sec:xu}

Notice that the code design in (\ref{eqn:code}) can only achieve the full diversity with PIC group decoding for two diagonal layers (r.f. \emph{Theorem 2}) and the code rate is not larger than $2$ symbols per channel use. With  $P$ ($P>2$) diagonal layers in the code (\ref{eqn:code}), the rate can be increased but the independence among $P$ channel groups $\{\begin{array}{cccc}
                       \mathbf{G}_1  &  \mathbf{G}_2 &  \cdots &   \mathbf{G}_P
                     \end{array}
   \}$ in (\ref{eqn:H0}) is not satisfied, thereby may lose the full diversity gain.  In this section, we propose a new code design    which can achieve full diversity with PIC group decoding and a rate above $2$. 

\subsection{Code Design}

For $M=3p$ ($p$ is an integer), our proposed STBC $\mathbf{C}_M$ for $M$ transmit antennas is given by
\begin{equation}\label{eqn:codeword}
\mathbf{C}_{3p}=\left[
\begin{array}{cccccccccc}
 X_{1,1} &0&0 &0&0&0  &\cdots &0&0&0\\
 0& X_{2,2}&0 &0&0&0  &\ddots  &0&0&0\\
 0&0& X_{3,3} &0&0&0  &\ddots  &0&0&0\\
 X_{2,1}&0&0& X_{1,4}&0&0  &\ddots  &0&0&0\\
 0&X_{3,2}&0&0& X_{2,5}&0  &\ddots  &0&0&0\\
 0&0&X_{1,3}&0&0& X_{3,6} &\ddots &0&0&0\\
 \ddots &\ddots &\ddots &\ddots &\ddots &\ddots &\ddots &\ddots &\ddots &\ddots\\
 0 &0&0&0&0&0 &\ddots &X_{1,3p-2}&0&0\\
 0&0&0&0& 0&0 &\ddots &0&X_{2,3p-1}&0\\
 0&0&0&0& 0&0 &\ddots &0&0&X_{3,3p}\\
 0 &X_{1,2}&0&0&0&0 &\ddots &X_{2,3p-2}&0&0\\
 0&0&X_{2,3}&0& 0&0 &\ddots &0&X_{3,3p-1}&0\\
 X_{3,1}&0&0&0& 0&0 &\ddots &0&0&X_{1,3p}\\
 0&0&0&X_{3,4}&X_{1,5}&X_{2,6}&\ddots &0&0&0\\
 \vdots&\vdots&\vdots&\vdots&\vdots&\vdots&\ddots&\vdots&\vdots&\vdots\\
 0&0&0&0&0&0&\ddots &X_{3,3p-2}&X_{1,3p-1}&X_{2,3p}
\end{array}
\right],
\end{equation}
where the symbol vector $\mathbf{X}_i=[X_{i,1}, X_{i,2},\cdots,
X_{i,M}]^t$ is given by
$$\mathbf{X}_i=\mathbf{\Theta} \mathbf{s}_i, i=1,2,3,$$
$\mathbf{\Theta}$ is an $M\times M$ constellation rotation matrix given by (\ref{eqn:cyclo}) and
$\mathbf{s}_i=[s_{(i-1)M+1},s_{(i-1)M+2},\cdots,s_{iM}]^t$ is the
$M\times1$ information symbol vector.

For $M=3p-1$, our proposed STBC $\mathbf{C}_M$ is given by
\begin{equation*}\label{eqn:codeword2}
\mathbf{C}_{3p-1}=\left[
\begin{array}{cccccccccccc}
 X_{1,1} &0&0 &0&0&0  &\cdots &0&0&0&0&0\\
 0& X_{2,2}&0 &0&0&0  &\ddots  &0&0&0&0&0\\
 0&0& X_{3,3} &0&0&0  &\ddots  &0&0&0&0&0\\
 X_{2,1}&0&0& X_{1,4}&0&0  &\ddots  &0&0&0&0&0\\
 0&X_{3,2}&0&0& X_{2,5}&0  &\ddots  &0&0&0&0&0\\
 0&0&X_{1,3}&0&0& X_{3,6} &\ddots &0&0&0&0&0\\
 \ddots &\ddots &\ddots &\ddots &\ddots &\ddots &\ddots &\ddots &\ddots&\ddots &\ddots &\ddots\\
 0 &0&0&0&0&0 &\ddots &X_{1,3p-5}&0&0&0&0\\
 0&0&0&0& 0&0 &\ddots &0&X_{2,3p-4}&0&0&0\\
 0&0&0&0& 0&0 &\ddots &0&0&X_{3,3p-3}&0&0\\
 0&0&0&0& 0&0 &\ddots &X_{2,3p-5}&0&0&X_{1,3p-2}&0\\
 0&0&0&0& 0&0 &\ddots &0&X_{3,3p-4}&0&0&X_{2,3p-1}\\
 0&0&0&0& 0&0 &\ddots &0&0&X_{1,3p-3}&X_{3,3p-2}&0\\
 0 &X_{1,2}&0&0&0&0 &\ddots &0&0&0&X_{2,3p-2}&0\\
 0&0&X_{2,3}&0& 0&0 &\ddots &0&0&0&0&X_{3,3p-1}\\
 X_{3,1}&0&0&0& 0&0 &\ddots &0&0&0&0&X_{1,3p-1}\\
 0&0&0&X_{3,4}&X_{1,5}&X_{2,6}&\ddots &0&0&0&0&0\\
 \vdots&\vdots&\vdots&\vdots&\vdots&\vdots&\ddots&\vdots&\vdots&\vdots&\vdots&\vdots\\
 0&0&0&0&0&0&\ddots &X_{3,3p-5}&X_{1,3p-4}&X_{2,3p-3}&0&0
\end{array}
\right].
\end{equation*}

For $M=3p+1$, our proposed STBC $\mathbf{C}_M$ is given by
\begin{eqnarray}\label{eqn:codeword3}
\mathbf{C}_{3p+1}=\left[
\begin{array}{ccccccccccc}
 X_{1,1} &0&0 &0&0&0  &\cdots &0&0&0&0\\
 0& X_{2,2}&0 &0&0&0  &\ddots  &0&0&0&0\\
 0&0& X_{3,3} &0&0&0  &\ddots  &0&0&0&0\\
 X_{2,1}&0&0& X_{1,4}&0&0  &\ddots  &0&0&0&0\\
 0&X_{3,2}&0&0& X_{2,5}&0  &\ddots  &0&0&0&0\\
 0&0&X_{1,3}&0&0& X_{3,6} &\ddots &0&0&0&0\\
 \ddots &\ddots &\ddots &\ddots &\ddots &\ddots&\ddots &\ddots &\ddots &\ddots &\ddots\\
 0 &0&0&0&0&0 &\ddots &X_{1,3p-2}&0&0&0\\
 0&0&0&0& 0&0 &\ddots &0&X_{2,3p-1}&0&0\\
 0&0&0&0& 0&0 &\ddots &0&0&X_{3,3p}&0\\
 0 &0&0&0&0&0 &\ddots &X_{2,3p-2}&0&0&X_{1,3p+1}\\
 0&0&0&0& 0&0 &\ddots &0&X_{3,3p-1}&0&X_{2,3p+1}\\
 0&0&0&0& 0&0 &\ddots &0&0&X_{1,3p}&X_{3,3p+1}\\
 X_{3,1}&X_{1,2}&X_{2,3}&0& 0&0&\ddots&0&0& 0&0\\
 0&0&0&X_{3,4}&X_{1,5}&X_{2,6}&\ddots &0&0&0&0\\
 \vdots&\vdots&\vdots&\vdots&\vdots&\vdots&\ddots&\vdots&\vdots&\vdots\vdots\\
 0&0&0&0&0&0&\ddots &X_{3,3p-2}&X_{1,3p-1}&X_{2,3p}&0
\end{array}
\right].
\end{eqnarray}

\vspace*{3mm}
\begin{proposition}
The proposed STBC has the rate as follows:
\begin{eqnarray}\label{eqn:rate2}
R=\left\{\begin{aligned}
&\frac{9M}{4M+6}  \qquad  \mathrm{for} \quad M=3p;&\\
&\frac{9M}{4M+4}  \qquad  \mathrm{for} \quad M=3p-1;&\\
&\frac{9M}{4M+5}  \qquad  \mathrm{for} \quad M=3p+1.&
\end{aligned}\right.
\end{eqnarray}
\end{proposition}

\vspace*{1mm}
\begin{proof}
For $M=3p$, in the codeword (\ref{eqn:codeword}) $3M$ information symbols are sent over $M+3+\frac{M}{3}-1$ time slots. Thus, the code rate is $\frac{9M}{4M+6}$. Similarly, it is easy to prove the code rate for cases of  $M=3p-1$ and $M=3p+1$.
\end{proof}

\vspace*{3mm}\begin{remark} [Asymptotic Rate]
It is obvious that the code rate (\ref{eqn:rate2}) approaches to $9/4$   when a large number of transmit antennas are used. Its full diversity property will be proved in the next subsection. However, the full diversity code proposed by the first design in (\ref{eqn:code}) cannot achieve a rate more than 2, which was shown in \emph{Theorem 2}.
\end{remark}

\vspace*{3mm}\begin{remark} [Rate Comparison]
Note that the code design in (\ref{eqn:code}) can  achieve full diversity with the PIC group decoding for $P=2$ only and the rate is   $\frac{2M}{M+1}$. In Table I, the comparison of the code rate between the first code design in (\ref{eqn:code}) and the second design in (\ref{eqn:codeword})-(\ref{eqn:codeword3}) is given.
\end{remark}

\begin{table}[h]
\begin{center}\caption{Comparison in code rate $R$ (symbols per channel use)}
\begin{tabular}{|c|c|c|}
  \hline
  \begin{tabular}{c}
     \\
    \\\hline
    $M=2$  \\\hline
    $M=3$ \\\hline
    $M=4$\\\hline
    $M=5$\\\hline
    $M=6$\\\hline
    $M=7$\\\hline
    $M=8$\\
  \end{tabular} & \begin{tabular}{c}
    Code  (\ref{eqn:code}) \\
    \hline\hline
    \begin{tabular}{c}
      $P=2$    \\
            \hline
       4/3    \\
      \hline
        3/2  \\
        \hline
       8/5    \\
        \hline
        5/3  \\\hline
       12/7   \\\hline
        7/4 \\\hline
        16/9 \\
    \end{tabular} \\
  \end{tabular}
  &  \begin{tabular}{c}
    Code (\ref{eqn:codeword})-(\ref{eqn:codeword3}) \\
    \hline\hline
    \begin{tabular}{c}
       $P=3$\\
            \hline
        3/2 \\
      \hline
         3/2\\
      \hline
         12/7\\
      \hline
         15/8\\
      \hline
       9/5\\
      \hline
       21/11\\
       \hline
       2\\
    \end{tabular} \\
  \end{tabular} \\
  \hline
\end{tabular}
\end{center}
\end{table}

\vspace*{3mm}\begin{remark} [Decoding Complexity]
The decoding complexity of the proposed STBC with the PIC
group decoding is equivalent to the ML decoding of $M$ independent
information symbols jointly.  According to \emph{Definition 2}, the ML decoding complexity in the PIC group decoding algorithm is   $\mathcal{O}= 3 |\mathcal{A}|^{M}$.
\end{remark}

\subsection{Full Diversity with PIC Group Decoding}
Next, we show  that the
proposed STBC in (\ref{eqn:codeword}) achieves full diversity when a PIC group decoding is
used at the receiver.

\vspace*{3mm}
\begin{theorem}
Let the STBC $\mathbf{C}$ as described in (\ref{eqn:codeword}) be used
at the transmitter. There are $M$ transmit antennas and $1$ receive antenna.
If the received signal is decoded using the PIC group decoding with
the grouping scheme
$\mathcal{I}=\{\mathcal{I}_1,\mathcal{I}_2,\mathcal{I}_3\}$, where
$\mathcal{I}_i=\{(i-1)M+1,(i-1)M+2,\cdots,iM\}$ for $i=1,2,3$, then
the code $\mathbf{C}$ achieves the full diversity.
\end{theorem}
\vspace*{3mm}

In order to prove the \emph{Theorem 4}, let us first introduce the
following lemma.

\vspace*{3mm}
\begin{lemma}
Consider the system as described in \emph{Theorem 4} with $N=1$,
and the channel matrix $\mathbf{H}=[\begin{array}{cccc}
                                      h_1 & h_2 & \cdots & h_M
                                    \end{array}]^t$. Let
$\mathbf{g}_i=h_i\mathbf{\Theta}_i$, where $\mathbf{\Theta}_i$ denotes the $i$th row of the rotation matrix $\mathbf{\Theta}$ for
$i=1,2,\cdots,M$. Then, the equivalent channel matrix is given by
\begin{equation}\label{equi}
\mathcal{H}=[\begin{array}{ccc}
               \mathbf{G}_1  & \mathbf{G}_2 & \mathbf{G}_3
             \end{array}]=\left[
\begin{array}{ccc}
 \mathbf{g}_1&0&0\\
 0&\mathbf{g}_2&0\\
 0&0&\mathbf{g}_3\\
 \mathbf{g}_4&\mathbf{g}_1&0\\
 0&\mathbf{g}_5&\mathbf{g}_2\\
 \mathbf{g}_3&0&\mathbf{g}_6\\
 \vdots&\vdots&\vdots\\
 \mathbf{g}_{3p-2}&\mathbf{g}_{3p-5}&0\\
 0&\mathbf{g}_{3p-1}&\mathbf{g}_{3p-4}\\
 \mathbf{g}_{3p-3}&0&\mathbf{g}_{3p}\\
 \mathbf{g}_2&\mathbf{g}_{3p-2}&0\\
 0&\mathbf{g}_3&\mathbf{g}_{3p-1}\\
 \mathbf{g}_{3p}&0&\mathbf{g}_1\\
 \mathbf{g}_5&\mathbf{g}_6&\mathbf{g}_4\\
 \vdots&\vdots&\vdots\\
 \mathbf{g}_{3p-1}&\mathbf{g}_{3p}&\mathbf{g}_{3p-2}
\end{array}
\right].
\end{equation}
\end{lemma}
\vspace*{3mm}

The proof of \emph{Lemma 3} is straightforward and it is easy to verify that
$\mathbf{C}\mathbf{H}=\mathcal{H}[\begin{array}{ccc}
                                    \mathbf{s}_1^t  & \mathbf{s}_2^t & \mathbf{s}_3^t
                                  \end{array}]^t$.

\vspace*{3mm}
\begin{proof}[Proof of Theorem 4]

Firstly, we show that the proposed STBC $\mathbf{C}$ in
(\ref{eqn:codeword}) achieves the full diversity with ML decoding, i.e.,
$\Delta \mathbf{C}=\mathbf{C}-\mathbf{C}'$ achieves full rank for
any distinct pair of codewords $\mathbf{X}$ and $\mathbf{X}'$.

Since $\mathbf{X}$ and $\mathbf{X}'$ are distinct, at least one pair
of the information symbol vectors $\mathbf{s}_i$ and $\mathbf{s}'_i$
are different. Suppose only one pair of the information symbol
vectors are distinct, say $\mathbf{s}_1$ and $\mathbf{s}'_1$. By the
\emph{Property 1}, $\Delta X_{1,j}=X_{1,j}-X'_{1,j}\neq0$ for all
$j=1,2,\cdots, M$. In this case, the matrix $\Delta \mathbf{C}$ is
exactly a diagonal matrix with the rows rearranged. Thus, it
achieves full rank.

If all three pairs of the information symbol vectors are distinct,
then $\Delta X_{i,j}=X_{i,j}-X'_{i,j}\neq0$ for all $i=1,2,3$ and
$j=1,2,\cdots, M$. It is easy to see from the codeword structure that
$\Delta \mathbf{C}$ is exactly a lower triangular matrix with
nonzero diagonal entries. Thus, it achieves full rank.

If two pairs of the information symbol vectors are distinct and the
other one pair of the information symbol vectors are the same, say $\mathbf{s}_1=\mathbf{s}'_1$. After replacing
the ($3q-2$)th row by the ($3q+1$)th row for
$q=1,2,\cdots,\frac{M}{3}$, the matrix $\Delta \mathbf{C}$ becomes a lower triangular
matrix with nonzero diagonal entries. Thus, it achieves full rank.

As shown above, $\Delta \mathbf{C}$ has full rank for all
$[\begin{array}{ccc}
    \mathbf{s}_1   & \mathbf{s}_2  & \mathbf{s}_3
  \end{array}] \neq[\begin{array}{ccc}
                        \mathbf{s}'_1  & \mathbf{s}'_2   & \mathbf{s}'_3
                     \end{array}] $,
i.e., $\mathbf{X}\neq \mathbf{X}'$.

\vspace*{3mm}
Next, we show that $\mathbf{G}_1$, $\mathbf{G}_2$ and
$\mathbf{G}_3$ are linearly independent vector groups as long as
$\mathbf{H}\neq0$.

If $h_1\neq0$, it is obvious that $\mathbf{G}_1$ is linearly independent of
$\mathbf{G}_2$ and $\mathbf{G}_3$. Otherwise, $h_1=0$, if
$h_4\neq0$, $\mathbf{G}_1$ is also independent of $\mathbf{G}_2$ and
$\mathbf{G}_3$. Therefore, we can reorder $\mathbf{H}$ as
$$\{h_1,h_4,\cdots,h_{3p-2},h_{3p},h_{3p-3},\cdots,h_3,h_2,h_5,\cdots,h_{3p-1}\}.$$
Suppose $h_j$ is the first one in the sequence such that $h_j\neq0$.
According to the \emph{Property 2}, the entries of $\mathbf{g}_j$
in $\mathbf{G}_1$ are not zeros, but the entries in the same row of
$\mathbf{G}_2$ and $\mathbf{G}_3$ are either 0 or
$\mathbf{g}_i=h_i\mathbf{\Theta}_i$ with $h_i=0$, which implies that
$\mathbf{G}_1$ can not be expressed as the linear combination of
$\mathbf{G}_2$ and $\mathbf{G}_3$.

Similarly, in order to prove $\mathbf{G}_2$ and $\mathbf{G}_3$ are
independent of the remaining vector groups, we only need to
rearrange the channel matrix as
$$\{h_2,h_5,\cdots,h_{3p-1},h_{3p-2},h_{3p-5},\cdots,h_1,h_3,h_6,\cdots,h_{3p}\}$$
and
$$\{h_3,h_6,\cdots,h_{3p},h_{3p-1},h_{3p-4},\cdots,h_2,h_1,h_4,\cdots,h_{3p-2}\},$$
respectively. Hence, $\mathbf{G}_1$, $\mathbf{G}_2$ and
$\mathbf{G}_3$ are linearly independent vector groups as long as
$\mathbf{H}\neq0$.

According to \emph{Proposition 1}, the code $\mathbf{C}$ in
(\ref{eqn:codeword}) achieves full diversity using PIC grouping decoding
with the group scheme
$\mathcal{I}=\{\mathcal{I}_1,\mathcal{I}_2,\mathcal{I}_3\}$.
\end{proof}

Similarly, we can prove that the codes $\mathbf{C}_{3p-1}$ and $\mathbf{C}_{3p+1}$ achieve full diversity with PIC grouping
decoding.

\subsection{Code Design Examples}

For $M=4$. It has $T=7$.
\begin{eqnarray}
\mathbf{C}_{4}=\left[\begin{array}{cccc}
                   X_{11} & 0  & 0 & 0  \\
                   0  & X_{22} & 0  & 0  \\
                    0 & 0  &  X_{33} & 0 \\
                   X_{21} & 0  & 0  & X_{14} \\
                    0 & X_{32} &  0 & X_{24} \\
                    0 &  0 & X_{13} & X_{34} \\
                   X_{31} & X_{12} & X_{23} &0
                 \end{array}
\right],
 \end{eqnarray}
 The   rate of this code is $12/7$.
 The equivalent channel of the code $\mathbf{C}_{4}$ is
 \begin{eqnarray}
 \mathcal{H}=\left[\begin{array}{ccc}
               \mathbf{g}_1 &  \mathbf{0} &  \mathbf{0} \\
                \mathbf{0} & \mathbf{g}_2 &  \mathbf{0} \\
                \mathbf{0} &  \mathbf{0} & \mathbf{g}_3 \\
               \mathbf{g}_4 & \mathbf{g}_1 & \mathbf{0}  \\
                \mathbf{0} & \mathbf{g}_4 & \mathbf{g}_2 \\
               \mathbf{g}_3 &   \mathbf{0}& \mathbf{g}_4 \\
               \mathbf{g}_2 & \mathbf{g}_3 & \mathbf{g}_1
             \end{array}\right],
  \end{eqnarray}
where the $1\times 4$ row vector $\mathbf{g}_{i}=\mathbf{\theta}_ih_i$, $i=1,2,3,4$, with $\mathbf{\theta}_i$ being the $i$th row of the matrix $\mathbf{\Theta}$.
 The grouping scheme for  the PIC group decoding is $\mathcal{I}_1=\{1,2, 3, 4\}$, $\mathcal{I}_2=\{5,6, 7,8\}$ and $\mathcal{I}_3=\{9,10,11,12\}$. Obviously, the code can obtain the full diversity with the PIC group decoding.

 For $M=6$.   It has $T=10$.
\begin{eqnarray}
\mathbf{C}_{6}=\left[ \begin{array}{cccccc}
                                    X_{1,1} & 0  & 0  &0&0&0\\
                                     0 & X_{2,2} & 0   &0&0&0 \\
                                     0 & 0  & X_{3,3} &0&0&0\\
                                    X_{2,1} & 0  &  0 & X_{1,4} &  0 &  0 \\
                                    0  & X_{3,2} &  0 & 0  & X_{2,5} &   0\\
                                     0 &  0 & X_{1,3} &  0 &  0 & X_{3,6} \\
                                     0 & X_{1,2} &  0 & X_{2,4} &  0 &  0 \\
                                     0 &  0 & X_{2,3} &  0 & X_{3,5} &   0\\
                                    X_{3,1} & 0  &  0 &  0 & 0  & X_{1,6} \\
                                     0 &  0 &  0 & X_{3,4} & X_{1,5} & X_{2,6}
                                  \end{array}
\right],
 \end{eqnarray}
The   rate of this code is $18/10$.
 The equivalent channel of the code $\mathbf{C}_{6}$ is
 \begin{eqnarray}
 \mathcal{H}=\left[ \begin{array}{ccc}
                      \mathbf{g}_1 &\mathbf{0}   & \mathbf{0}  \\
                       \mathbf{0} & \mathbf{g}_2 & \mathbf{0}  \\
                       \mathbf{0} & \mathbf{0}  & \mathbf{g}_3  \\
                      \mathbf{g}_4  & \mathbf{g}_1  & \mathbf{0}  \\
                      \mathbf{0}  & \mathbf{g}_5  & \mathbf{g}_2  \\
                      \mathbf{g}_3  &  \mathbf{0} & \mathbf{g}_6  \\
                      \mathbf{g}_2  & \mathbf{g}_4  & \mathbf{0}  \\
                       \mathbf{0} & \mathbf{g}_3  & \mathbf{g}_5 \\
                      \mathbf{g}_6  & \mathbf{0}  & \mathbf{g}_1  \\
                      \mathbf{g}_5  & \mathbf{g}_6  & \mathbf{g}_4
                    \end{array}
 \right],
  \end{eqnarray}
where the $1\times 6$ row vector $\mathbf{g}_{i}=\mathbf{\theta}_ih_i$, $i=1,2,3,4,5,6$, with $\mathbf{\theta}_i$ being the $i$th row of the matrix $\mathbf{\Theta}$.
 The grouping scheme for  the PIC group decoding is $\mathcal{I}_1=\{1,2, 3, 4, 5,6\}$, $\mathcal{I}_2=\{7,8,9,10,11,12\}$ and $\mathcal{I}_3=\{13, 14, 15, 16, 17, 18\}$. Obviously, the code can obtain the full diversity with the PIC group decoding.

\vspace*{3mm}
A code example of $M=9$  is given by
\begin{equation*}
\mathbf{C}_9=\left[
\begin{array}{ccccccccc}
 X_{1,1} &0&0 &0&0&0  &0&0&0\\
 0& X_{2,2}&0 &0&0&0    &0&0&0\\
 0&0& X_{3,3} &0&0&0   &0&0&0\\
 X_{2,1}&0&0& X_{1,4}&0&0    &0&0&0\\
 0&X_{3,2}&0&0& X_{2,5}&0    &0&0&0\\
 0&0&X_{1,3}&0&0& X_{3,6}  &0&0&0\\
 0 &0&0&X_{2,4}&0&0  &X_{1,7}&0&0\\
 0&0&0&0&X_{3,5}&0 &0&X_{2,8}&0\\
 0&0&0&0& 0&X_{1,6}&0&0&X_{3,9}\\
 0 &X_{1,2}&0&0&0&0 &X_{2,7}&0&0\\
 0&0&X_{2,3}&0& 0&0 &0&X_{3,8}&0\\
 X_{3,1}&0&0&0& 0&0 &0&0&X_{1,9}\\
 0&0&0&X_{3,4}&X_{1,5}&X_{2,6} &0&0&0\\
 0&0&0&0&0&0&X_{3,7}&X_{1,8}&X_{2,9}
\end{array}
\right].
\end{equation*}


\section{Simulation Results}\label{sec:sim}

In this section, simulation results of the proposed STBC with the PIC group decoding over Rayleigh fading channels are presented.
 We first show   bit error rate (BER) performance of the codes  proposed  in this paper  for four transmit antennas and compare them to the one proposed in \cite{Guo}.
 Specifically, we consider   three STBC for four transmit antennas  proposed in this paper, i.e.,  $\mathbf{C}_{4,6,3}$ in (\ref{eqn:C463}), $\mathbf{C}_{4,6,2}$ in (\ref{eqn:C462}) and $\mathbf{C}_{4,5,2}$ in (\ref{eqn:C452}), and then compare them with  Guo-Xia's code given in \cite[Section VI - \emph{Example 2}]{Guo}. In order to make a fair performance comparison, we keep the same bandwidth efficiency of $8$ bps/Hz. Thus, we use 16QAM for the  code $\mathbf{C}_{4,6,3}$ (the code rate is $R=2$) and 64QAM for both codes $\mathbf{C}_{4,6,2}$ (the code rate  is $R=\frac{4}{3}$) and Guo-Xia's code (the code rate is $R=\frac{4}{3}$). For the code $\mathbf{C}_{4,5,2}$, because it has a code rate $R=\frac{8}{5}$ we  use 64QAM and thus its bandwidth efficiency is $9.6$ bps/Hz higher than the other three codes.
Since we use square QAM, in the rotation matrix $\mathbf{\Theta}$
we use $m=4$ in \emph{Example 1} of Section \ref{sec:new}.

\begin{figure}[t]
\centering
\includegraphics[width=5in]{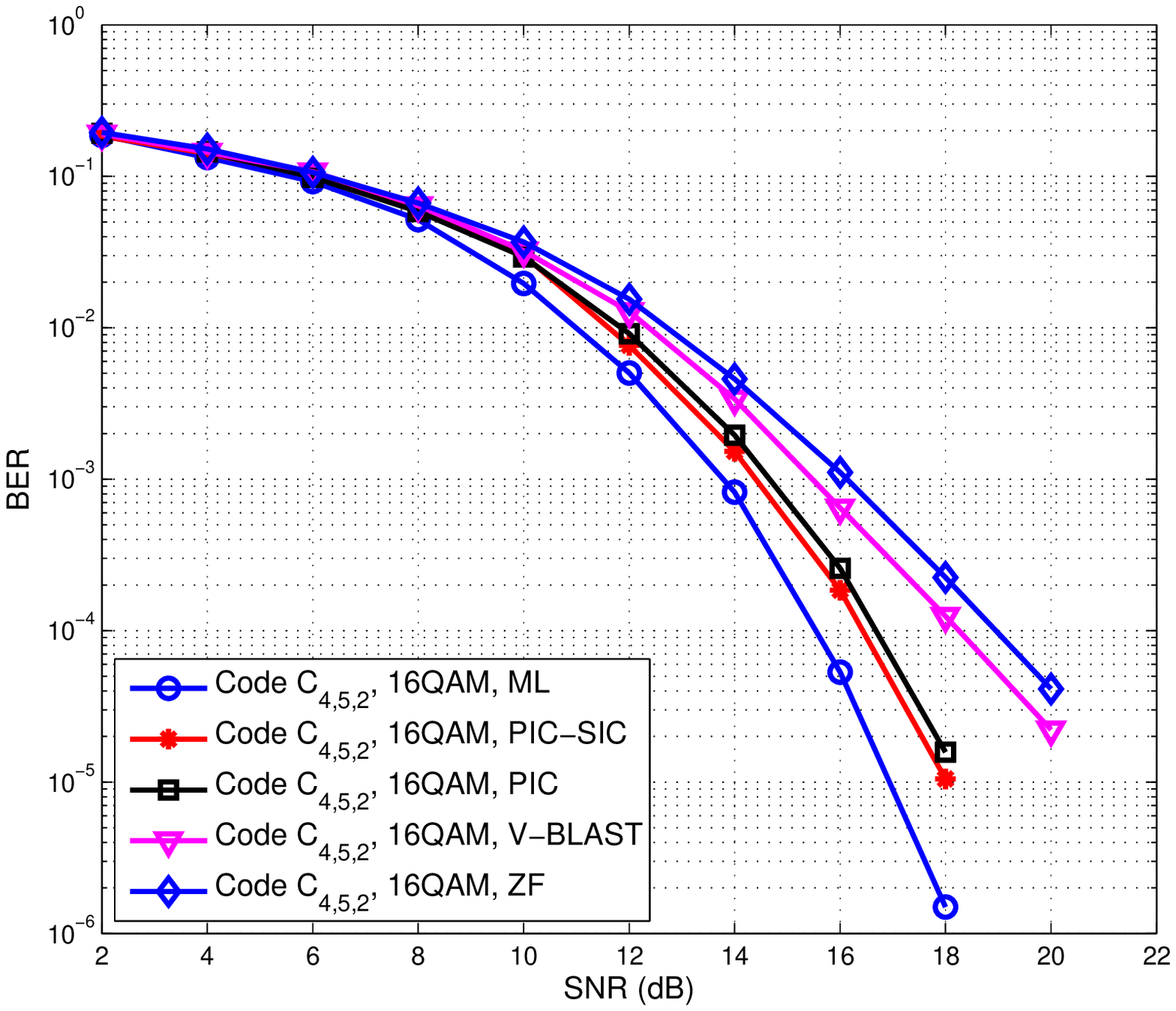}
\caption{BER performance of the proposed code $\mathbf{C}_{4,5,2}$ with various detection methods over $4\times 4$ Rayleigh fading channels.} \label{fig:C452}
\end{figure}

Fig. \ref{fig:C452} shows the performance of the proposed code $\mathbf{C}_{4,5,2}$ with various detection approaches for a $4\times 4$ MIMO system over Rayleigh fading channels. It can be seen that the code with PIC and PIC-SIC group decoding algorithms has both less than $1$ dB SNR performance loss compared to the one with the ML decoding. The performance loss of the PIC and PIC-SIC group decoding is trade for the largely reduced decoding complexity. Specifically, for 16QAM signaling the ML decoding complexity is $\mathcal{O}_{ML}=16^{8}=2^{32}$ and the complexity of the PIC group decoding is $\mathcal{O}_{PIC}=2\times 16^{4}=2^{17}$. Moreover, it is obvious that the code $\mathbf{C}_{4,5,2}$ with the PIC group decoding and the PIC-SIC group decoding can both obtain full diversity at high SNR. BLAST and ZF detection of the code $\mathbf{C}_{4,5,2}$ cannot obtain full diversity.

\begin{figure}[t]
\centering
\includegraphics[width=5in]{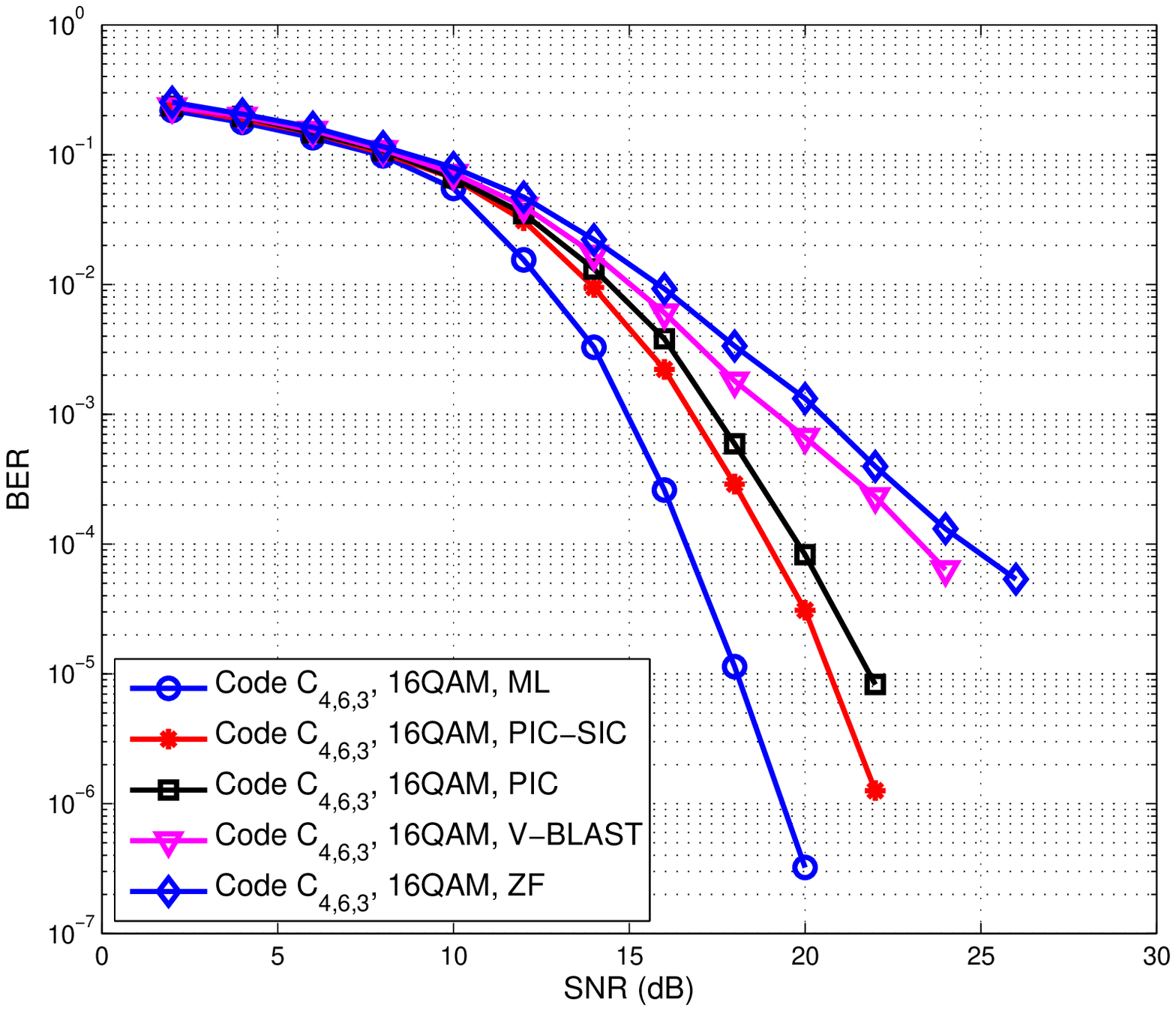}
\caption{BER performance of the proposed code $\mathbf{C}_{4,6,3}$ with various detection methods over $4\times 4$ Rayleigh fading channels.} \label{fig:C463}
\end{figure}

Fig. \ref{fig:C463} shows the performance of the proposed code $\mathbf{C}_{4,6,3}$ with various detection approaches for a $4\times 4$ MIMO system over Rayleigh fading channels.  It is shown that  the code $\mathbf{C}_{4,6,3}$ with PIC group decoding does not guarantee the full diversity at high SNR. However, with PIC-SIC group decoding the code $\mathbf{C}_{4,6,3}$ can achieve the full diversity. The performance gap between the ML decoding and the PIC  group decoding is around 2 dB SNR loss, but the decoding complexity is significantly reduced by the PIC group decoding from $16^{12}$ to $3\times 16^4$. Again, ZF and BLAST detection cannot obtain the full diversity.

\begin{figure}[t]
\centering
\includegraphics[width=5in]{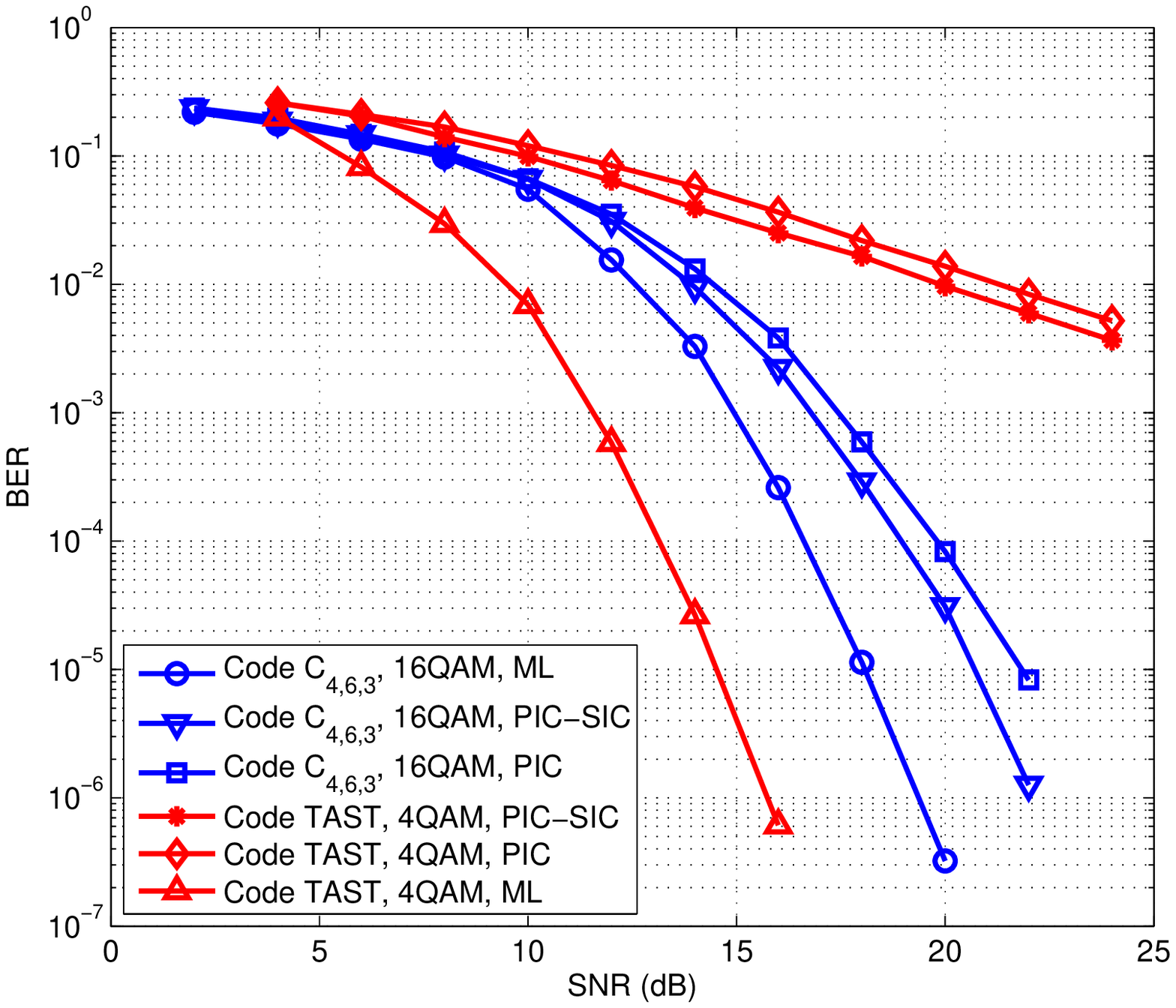}
\caption{Performance comparison between the proposed code $\mathbf{C}_{4,6,3}$  and the TAST code \cite{gamal}  for a MIMO system with  4 transmit antennas and 4 receive antennas at 8 bps/Hz.} \label{fig:TAST}
\end{figure}

Fig. \ref{fig:TAST} presents the performance comparison between  TAST code \cite{gamal} and the proposed code $\mathbf{C}_{4,6,3}$ with ML, PIC and PIC-SIC group decoding algorithms, respectively, for $4\times 4$ systems over Rayleigh fading channels. Note the   rate of TAST code is full, i.e., $4$ symbols per channel use for $4$ transmit antennas. To keep the bandwidth efficiency of $8$ bps/Hz, 4-QAM is used for TAST code and 16-QAM is used for the code $\mathbf{C}_{4,6,3}$.  It is shown in Fig. \ref{fig:TAST} that TAST code with the ML decoding gives the best performance. However, with PIC and PIC-SIC group decoding algorithms   TAST code will lose full diversity. The proposed code $\mathbf{C}_{4,6,3}$ can obtain much better performance than  TAST code when both use the PIC and PIC-SIC group decoding algorithms.

\begin{figure}[t]
\centering
\includegraphics[width=5in]{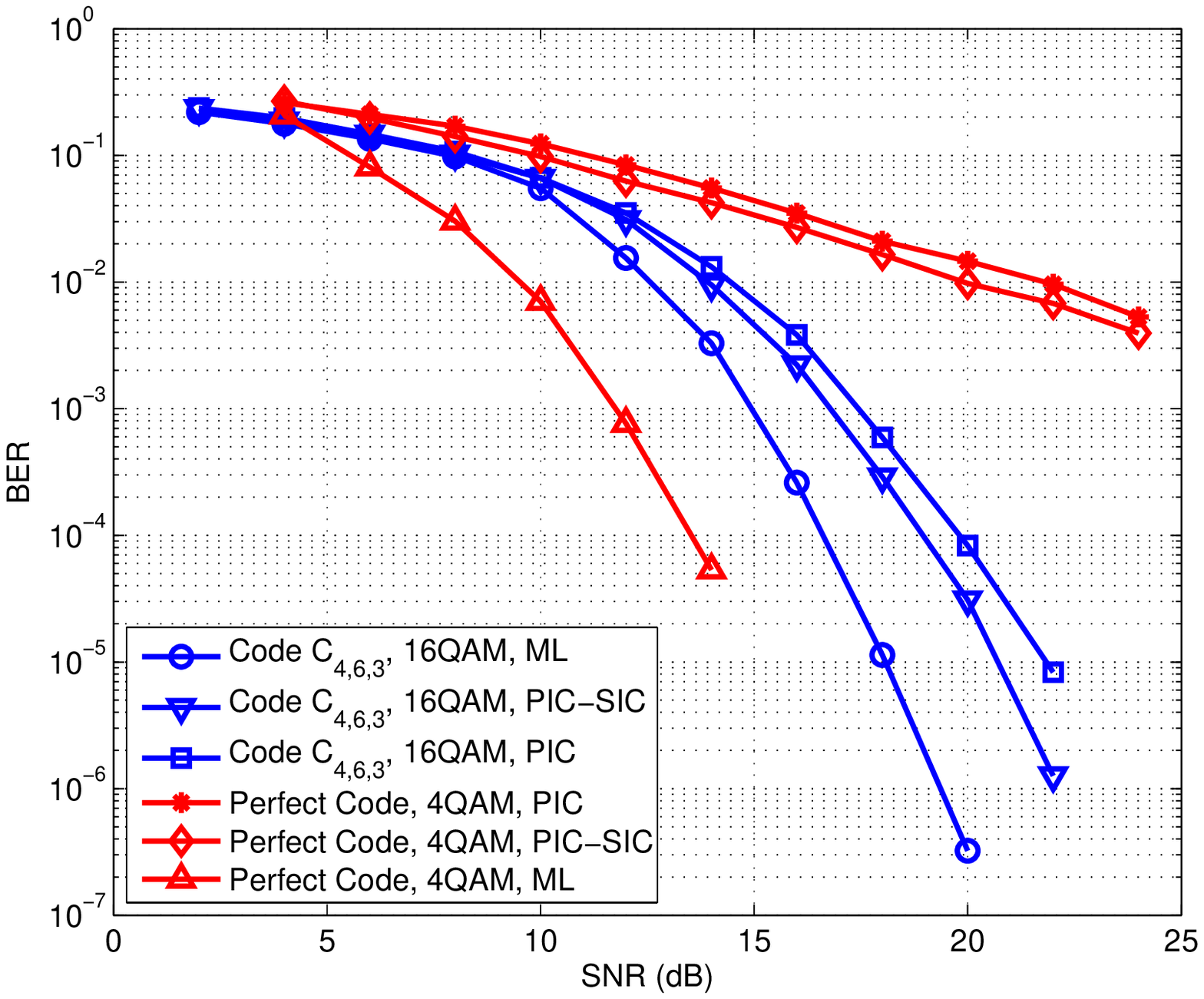}
\caption{Performance comparison between the proposed code $\mathbf{C}_{4,6,3}$  and the perfect ST code \cite{oggier}  for a MIMO system with  4 transmit antennas and 4 receive antennas at 8 bps/Hz.} \label{fig:perfect}
\end{figure}

Fig. \ref{fig:perfect} gives the performance comparison   between   Perfect STBC \cite{oggier} and the proposed code $\mathbf{C}_{4,6,3}$ with ML, PIC and PIC-SIC group decoding algorithms, respectively, for $4\times 4$ systems over Rayleigh fading channels. Similar to  TAST code,   Perfect STBC cannot obtain full diversity with the PIC and PIC-SIC group decoding. The proposed code $\mathbf{C}_{4,6,3}$  has a much better performance than Perfect STBC while admitting low complexity in the decoding.

\begin{figure}[t]
\centering
\includegraphics[width=5in]{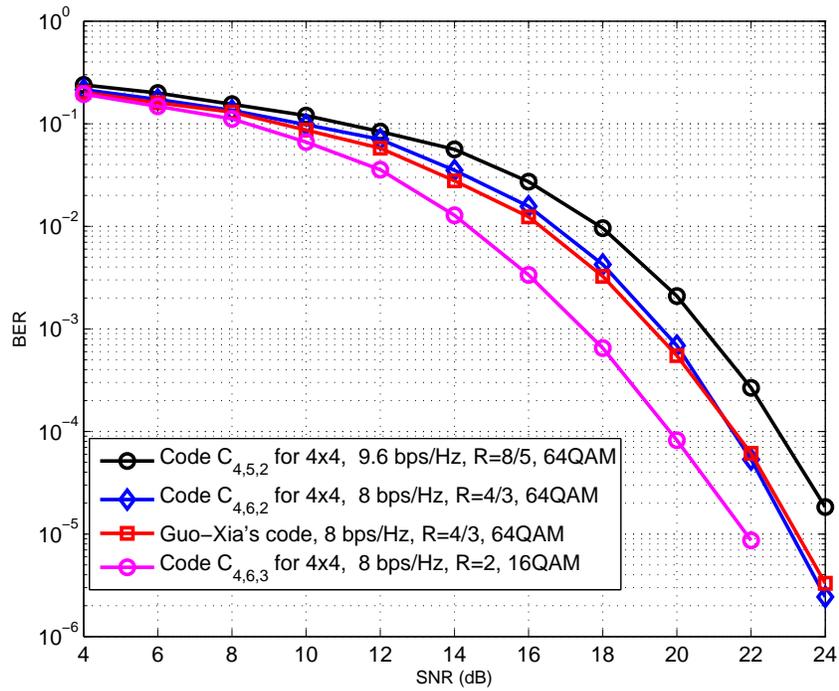}
\caption{BER performance of various codes with PIC group decoding for a MIMO system with  4 transmit antennas and 4 receive antennas.} \label{fig:Codes}
\end{figure}

Fig. \ref{fig:Codes} shows the BER performance of various codes with the PIC group decoding for $4$ transmit and $4$ receive antennas. It can be seen that the code   $\mathbf{C}_{4,6,2}$ has very similar performance to Guo-Xia's code. Moreover, the code  $\mathbf{C}_{4,5,2}$ has $1$ dB loss compared to Guo-Xia's code. This is because it has a higher bandwidth efficiency than Guo-Xia's code.
 In particular, the code $\mathbf{C}_{4,6,3}$ in (\ref{eqn:C463}) achieves the best BER performance among all the simulated codes. This is attributed to more information symbols embedded in the code $\mathbf{C}_{4,6,3}$ than the other codes and its code rate is the highest.

\begin{figure}[t]
\centering
\includegraphics[width=5in]{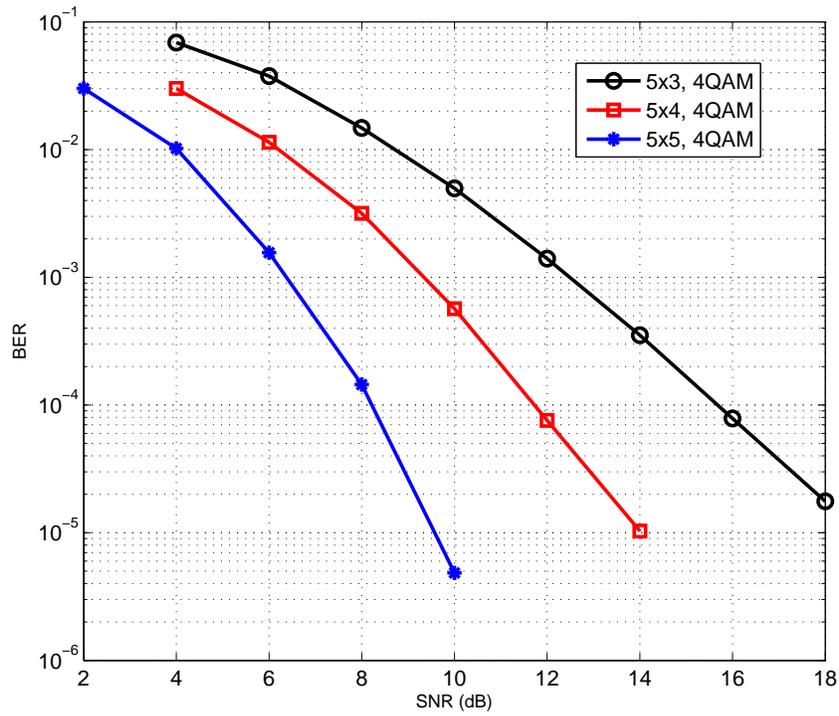}
\caption{BER performance of the proposed code  $\mathbf{C}_{5,6,2}$ with PIC group decoding for a MIMO system with 5 transmit antennas and different number of receive antennas.} \label{fig:Code5Tx}
\end{figure}

We also consider the simulation of the proposed code $\mathbf{C}_{5,6,2}$ in (\ref{eqn:C5}) with the PIC group decoding for five transmit antennas. Fig. \ref{fig:Code5Tx} shows the BER performance of the  code  $\mathbf{C}_{5,6,2}$ for three, four and five receive antennas, respectively. It demonstrates that an increase of the number of receive antennas results in a larger diversity gain as illustrated by the slope of the BER curves.

\section{Conclusion}\label{sec:conclusion}
In this paper, two designs of STBC that can achieve full diversity with PIC group decoding was proposed. The first proposed STBC are  constructed with multiple diagonal layers  and each layer is composed of a fixed number of coded symbols equal to $M$, i.e., the number of the transmit antennas. The code rate of the first proposed STBC is varied in accordance with the number of layers embedded in the codeword.  For the PIC group decoding, a grouping scheme was proposed to cluster every $M$ neighboring columns of the equivalent channel matrix  into one group. With the proposed STBC and the PIC group decoding in MIMO systems, it was proved that full diversity can be achieved when two diagonal layers are embedded in the code matrix.  Moreover,  for the full-diversity STBC with the PIC group decoding, the code rate is up to $2$ symbols per channel use.  When PIC-SIC group decoding is used, the proposed full-diversity STBC can have a rate up to $M$. A few examples of     code design achieving full diversity PIC group decoding   were  given.  The second proposed STBC
are based on three-layers and can achieve full diversity with the PIC group
decoding and their code rates can be up to $9/4$.
 Simulation results confirmed the theoretical analysis and show the full diversity performance of the proposed codes when the PIC group decoding is applied at receiver. It was also demonstrated that the proposed STBC  outperform Perfect STBC and TAST code when PIC group decoding is applied.


\section*{Appendix I - Proof of Lemma 1}

\subsection{Proof of Lemma 1.1}
 \begin{proof}
   Define the $T\times M$ matrix $\mathbf{C}_p$ ($p=1,2,\cdots, P$) as follows,
    \begin{eqnarray}\label{eqn:Cp}
\mathbf{C}_p = \left[\begin{array}{c}
                        \mathbf{0}_{(p-1)\times M} \\
                        \mathrm{diag}(\mathbf{X}_p) \\
                        \mathbf{0} _{(P-p)\times M}
                      \end{array}
 \right],\,\,\,p=1,2,\cdots,P,
 \end{eqnarray}
 where  $P=T-M+1$ and
 $\mathbf{X}_p =\left[\begin{array}{cccc}
                      X_{p,1}   & X_{p,2}& \cdots &  X_{p,M}
                    \end{array}
  \right]^t$ is given by (\ref{eqn:RM}).
 Then, (\ref{eqn:code}) can be  written as
       \begin{eqnarray}\label{eqn:code2}
 \mathbf{C}= \sum_{p=1}^P \mathbf{C}_p.
 \end{eqnarray}

  For MISO systems, we have $\mathbf{H}=\mathbf{h}=[\begin{array}{cccc}
                                                      h_1 & h_2 & \cdots & h_M
                                                    \end{array}
  ]^t$ with $h_m$ ($m=1,2\cdots, M$) being the channel gain from the $m$-th transmit antenna to the receiver. Using (\ref{eqn:code2}), we can express (\ref{eqn:Y}) as
   \begin{eqnarray}\label{eqn:Ya}
 \mathbf{y}&=&\sqrt{\frac{\rho}{\mu}}\sum_{p=1}^P \mathbf{C}_p \mathbf{h}+\mathbf{w}\nonumber\\
 &=&\sqrt{\frac{\rho}{\mu}}\sum_{p=1}^P \mathcal{H}_p \mathbf{X}_p+\mathbf{w},
 \end{eqnarray}
   where $\mathbf{y}\in\mathbb{C}^{T\times 1}$, $\mathbf{w}\in\mathbb{C}^{T\times 1}$  and  $\mathcal{H}_p\in\mathbb{C}^{T\times M}$ is given by
   \begin{eqnarray}\label{eqn:Hp}
\mathcal{H}_p = \left[\begin{array}{c}
                        \mathbf{0}_{(p-1)\times M} \\
                        \mathrm{diag}(\mathbf{h}) \\
                        \mathbf{0} _{(P-p)\times M}
                      \end{array}
 \right],\,\,\,p=1,2,\cdots,P.
 \end{eqnarray}
Using (\ref{eqn:RM}), we can further  write (\ref{eqn:Ya}) as
   \begin{eqnarray} \label{eqn:Y3}
 \mathbf{y} &=&\sqrt{\frac{\rho}{\mu}}\sum_{p=1}^P \mathcal{H}_p \mathbf{\Theta}\mathbf{s}_p+\mathbf{w}\nonumber\\
 &=&\sqrt{\frac{\rho}{\mu}}\mathcal{H}\mathbf{s}+\mathbf{w},
 \end{eqnarray}
 where the equivalent channel matrix $\mathcal{H}\in\mathbb{C}^{T\times MP}$   is given by
  \begin{eqnarray}\label{eqn:eqH}
   \mathcal{H}=\left[\begin{array}{cccc}
                      ( \mathcal{H}_1 \mathbf{\Theta})  &  (\mathcal{H}_2 \mathbf{\Theta}) &  \cdots &   (\mathcal{H}_P \mathbf{\Theta})
                     \end{array}
   \right]
  \end{eqnarray}
 and $\mathbf{s}=[\begin{array}{cccc}
                      \mathbf{s}_1^t & \mathbf{s}_2^t & \cdots & \mathbf{s}_{P}^t
                    \end{array}
   ]^t$.

 Let $\mathbf{G}_p=\mathcal{H}_p \mathbf{\Theta}$  for $p=1,2,\cdots, P$. Using (\ref{eqn:Hp}),  we get
   \begin{eqnarray}
   \mathbf{G}_p= \left[\begin{array}{c}
                        \mathbf{0}_{(p-1)\times M} \\
                        \mathrm{diag}(\mathbf{h})\mathbf{\Theta} \\
                        \mathbf{0} _{(P-p)\times M}
                      \end{array}
 \right],\,\,\,\,\,p=1,2,\cdots, P.
   \end{eqnarray}
   Then, (\ref{eqn:eqH}) can be written as
\begin{eqnarray}
   \mathcal{H}=\left[\begin{array}{cccc}
                       \mathbf{G}_1  &  \mathbf{G}_2 &  \cdots &   \mathbf{G}_P
                     \end{array}
   \right].
  \end{eqnarray}
\end{proof}

\subsection{Proof of Lemma 1.2}
\begin{proof}
   Next, we shall prove   that $\mathbf{G}_1$ and $\mathbf{G}_2$    are linearly independent vector groups as long as $\mathbf{h}\neq \mathbf{0}$. To do so, we may need the following definitions.

\vspace*{3mm}

   \begin{definition} \label{def1}\cite{Guo}. Let $\mathcal{V}=\{\mathbf{v}_i\in \mathbb{C}^n, i=0, 1,\cdots, k-1\}$ be a set of vectors. Vector $\mathbf{v}_k$ is said to be independent of $\mathcal{V}$ if for any $a_i\in \mathbb{C},  i=0, 1,\cdots, k-1$,
   $$
   \mathbf{v}_k-\sum_{i=0}^{k-1}a_i\mathbf{v}_i\neq \mathbf{0}.
   $$
\end{definition}

\vspace*{3mm}

     \begin{definition} \label{def2}\cite{Guo}. Let $\mathcal{V}_0, \mathcal{V}_1,\cdots, \mathcal{V}_{n-1}, \mathcal{V}_{n}$ be $n+1$ groups of vectors. Vector group $\mathcal{V}_n$ is said to be independent of $\mathcal{V}_0, \mathcal{V}_1,\cdots, \mathcal{V}_{n-1}$ if every vector in $\mathcal{V}_n$ is independent of $\bigcup_{i=0}^{n-1}\mathcal{V}_i$.
\end{definition}

\vspace*{3mm}

      \begin{definition} \label{def3}\cite{Guo}. Let $\mathcal{V}_0, \mathcal{V}_1,\cdots, \mathcal{V}_{n-1}, \mathcal{V}_{n}$ be $n+1$ groups of vectors. The vector groups $\mathcal{V}_0, \mathcal{V}_1,\cdots, \mathcal{V}_{n-1}, \mathcal{V}_{n}$ are said to be linearly independent if for $0\leq k\leq n$, $\mathcal{V}_k$ is independent of the remaining vector groups $\mathcal{V}_0, \mathcal{V}_1,\cdots,\mathcal{V}_{k-1},  \mathcal{V}_{k+1}, \cdots, \mathcal{V}_{n}$.
\end{definition}

\vspace*{3mm}

To prove that $\mathbf{G}_1$ and $\mathbf{G}_2$ are linearly independent, from \emph{Definition \ref{def3}} it is equivalent to prove the  following two steps:
\begin{enumerate}
  \item  Prove that $\mathbf{G}_1$ is independent of $\mathbf{G}_2$ for any $\mathbf{h}\neq \mathbf{0}$.
  \item  Prove that $\mathbf{G}_2$ is independent of   $\mathbf{G}_{1}$ for any $\mathbf{h}\neq \mathbf{0}$.
\end{enumerate}
Using (\ref{eqn:G}), we can express $\mathbf{G}_1$ and $\mathbf{G}_2$ as, respectively,
 \begin{eqnarray}
   \mathbf{G}_1= \left[\begin{array}{c}
                        \mathrm{diag}(\mathbf{h})\mathbf{\Theta} \\
                        \mathbf{0} _{1\times M}
                      \end{array}
 \right],\,\,\,\,\,\,\,   \mathbf{G}_2= \left[\begin{array}{c}
                        \mathbf{0}_{1\times M} \\
                        \mathrm{diag}(\mathbf{h})\mathbf{\Theta}
                      \end{array}
 \right].
   \end{eqnarray}

\vspace*{3mm}

   \emph{Step 1} - $\mathbf{G}_1$ is independent of $\mathbf{G}_2$ for any $\mathbf{h}\neq \mathbf{0}$.

\vspace*{3mm}

For any $\mathbf{h}\neq \mathbf{0}$, we can find a  minimal index $\sigma$ ($1\leq \sigma\leq M$) such that $h_\sigma\neq 0$. That is, $h_1=\cdots=h_{\sigma-1}=0$.

  Let $\mathbf{g}_{1,m}$  be the $m$-th ($1\leq m\leq M$) column   of the matrix $\mathbf{G}_1$.  In order to prove that $\mathbf{G}_1$ is independent of $\mathbf{G}_2$, from \emph{Definition \ref{def2}} we see it is equivalent to prove that  the vector $\mathbf{g}_{1,m}$ is independent of     $\mathbf{G}_2$ for all $m=1,2,\cdots, M$. Further, using  \emph{Definition \ref{def1}} it is equivalent to prove that
   \begin{eqnarray}\label{eqn:G1}
   \mathbf{g}_{1,m}-  \mathbf{G}_2 \bar{\beta}_2\neq \mathbf{0}, \,\,\,\,\forall m, m=1,2,\cdots, M,
   \end{eqnarray}
   where $\bar{\beta}_2$ denotes an $M\times 1$ vector. Equivalently,  (\ref{eqn:G1}) can be expressed as
   \begin{eqnarray}\label{eqn:G1b}
   a\mathbf{g}_{1,m}- \mathbf{G}_2 a\bar{\beta}_2\neq \mathbf{0}, \,\,\,\,\forall m, m=1,2,\cdots, M,
   \end{eqnarray}
   where $a$ is a constant and $a\neq 0$.

   In order to prove (\ref{eqn:G1b}), we can use proof by contradiction. That is, we assume that
     $
   a\mathbf{g}_{1,m}- \mathbf{G}_2 a\bar{\beta}_2= \mathbf{0}, \,\,\,\,\forall m, m=1,2,\cdots, M,
   $
   and $a\neq 0$. Then, we examine the $\sigma$-th equation (from top to bottom) of (\ref{eqn:G1b}) and get $a{g}_{\sigma,m}^{(1)}=0$ for  $1\leq m\leq M$, where $g_{\sigma,m}^{(1)}$  denotes the $(\sigma,m)$-th entry of the matrix $\mathbf{G}_1$.
    This is because all top $\sigma$ rows of $\mathbf{G}_2$   are all zeros when $h_1=\cdots=h_{\sigma-1}=0$ as seen from (\ref{eqn:G}).
    Again from (\ref{eqn:G}), we have $g_{\sigma,m}^{(1)}=h_\sigma\theta_{\sigma,m}$ where $\theta_{\sigma,m}$ denotes the $(\sigma,m)$-th entry of the matrix $\mathbf{\Theta}$.
    For given $h_{\sigma}\neq0$ and  $\theta_{\sigma,m}\neq 0$ for all $m=1,2,\cdots,M$, we  get $a=0$. This contradicts with the assumption $a\neq 0$. Therefore, (\ref{eqn:G1b}) holds and \emph{Step 1} is proved.

\vspace*{3mm}
   \emph{Step 2} - $\mathbf{G}_2$ is independent of  $\mathbf{G}_{1}$ for any $\mathbf{h}\neq \mathbf{0}$.
\vspace*{3mm}

For any $\mathbf{h}\neq \mathbf{0}$, we can have a  maximal index $\upsilon$ ($1\leq \upsilon\leq M$) such that $h_\upsilon\neq 0$. That is, $h_{\upsilon+1}=\cdots=h_{M}=0$.

   Following the similar way in the proof of \emph{Step 1}, we can prove \emph{Step 2} as follows.

    Let $\mathbf{g}_{2,m}$ denote the $m$-th ($1\leq m\leq M$) column of $\mathbf{G}_2$. In order to prove that $\mathbf{G}_2$ is independent of $\mathbf{G}_{1}$, from  \emph{Definition \ref{def2}} it is equivalent to prove that  the vector $\mathbf{g}_{2,m}$ is independent of     $\mathbf{G}_1$ for all $m=1,2,\cdots, M$. Further, using \emph{Definition \ref{def1}} it is equivalent to prove that
   \begin{eqnarray}\label{eqn:Gq}
   \mathbf{g}_{2,m}-  \mathbf{G}_1 \bar{\beta}_1\neq \mathbf{0}, \,\,\,\,\forall m, m=1,2,\cdots, M,
   \end{eqnarray}
   where $\bar{\beta}_1$ denotes an $M\times 1$ vector. Equivalently,  (\ref{eqn:Gq}) can be expressed as
   \begin{eqnarray}\label{eqn:Gqb}
   a\mathbf{g}_{2,m}- \mathbf{G}_1 a \bar{\beta}_1\neq \mathbf{0}, \,\,\,\,\forall m, m=1,2,\cdots, M,
   \end{eqnarray}
   where $a$ is a constant and $a\neq 0$.

    In order to prove (\ref{eqn:Gqb}), we can use proof by contradiction. We assume that
     $
   a\mathbf{g}_{2,m}- \mathbf{G}_1 a\bar{\beta}_1= \mathbf{0}, \,\,\,\,\forall m, m=1,2,\cdots, M,
   $
   and $a\neq 0$. Note the first row of $\mathbf{G}_2$ is all-zero. Then, we examine the $(\upsilon+1)$-th equation(from top to bottom) of (\ref{eqn:Gqb}) and get $ag_{\upsilon+1,m}^{(2)}=0$ for  $1\leq m\leq M$, where $g_{\upsilon+1,m}^{(2)}$  denotes the $(\upsilon+1,m)$-th entry of the matrix $\mathbf{G}_2$. This is because the   $(\upsilon+1)$-th  row  of $\mathbf{G}_{1}$  is all-zero  when $h_{\upsilon+1}=\cdots=h_{M}=0$  as seen from (\ref{eqn:G}). Again from (\ref{eqn:G}), we have $g_{\upsilon+1,m}^{(2)}=h_\upsilon\theta_{\upsilon,m}$ where $\theta_{\upsilon,m}$ denotes the $(\upsilon,m)$-th entry of the matrix $\mathbf{\Theta}$.
    For given $h_{\upsilon}\neq0$ and  $\theta_{\upsilon,m}\neq 0$ for all $m=1,2,\cdots,M$, we then get $a=0$. This  contradicts with the assumption $a\neq 0$. Therefore, (\ref{eqn:Gqb}) holds and \emph{Step 2} is proved.

\vspace*{3mm}
   To summarize \emph{Step 1} and \emph{Step 2}, we prove that $\mathbf{G}_1$ and $\mathbf{G}_2$ are linearly independent vector groups for any $\mathbf{h}\neq \mathbf{0}$.
\end{proof}





%

 \newpage

\end{document}